\documentclass[journal]{IEEEtran}
\usepackage{multirow}
\usepackage{graphicx}
\usepackage[ruled,vlined]{algorithm2e}
\usepackage{booktabs}
\usepackage{color}
\newcommand{\PK}{\color{black}}
\usepackage{array}
\usepackage{pstool}
\usepackage{comment}

\ifCLASSINFOpdf
\else
  \usepackage[dvips]{graphicx}
   \DeclareGraphicsExtensions{.pdf}
\fi
\usepackage{amsmath}
\usepackage{tabu}
\usepackage{siunitx}[=v2]
\begin{document}
\newcolumntype{M}[1]{>{\centering\arraybackslash}m{#1}}

\tabulinesep=1.2mm

%
\title{Path Loss and Shadowing Modeling for Vehicle-to-Vehicle Communications in Terrestrial TV Band}
%
%
%

\author{Pawel~Kryszkiewicz,~\IEEEmembership{Senior Member,~IEEE,}
        Pawel~Sroka,~\IEEEmembership{Member,~IEEE,}
        Michal~Sybis,~\IEEEmembership{Member,~IEEE,}
        and~Adrian~Kliks,~\IEEEmembership{Senior Member,~IEEE,}
\thanks{All authors are with Institute of Radiocommunications, Poznan University of Technology, Poland e-mail: pawel.kryszkiewicz@put.poznan.pl.

The final version of record is available at
https://doi.org/10.1109/TAP.2022.3216472
}
}

%
%

\markboth{Submitted to IEEE Transactions on Antennas and Propagation}
{Kryszkiewicz \MakeLowercase{\textit{et al.}}: Path Loss and Shadowing Modeling for Vehicle-to-Vehicle Communications in Terrestrial TV Band}
%



\maketitle

\begin{abstract}
Vehicle platooning is considered as one of the key use cases for vehicle-to-vehicle (V2V) communications. However, its benefits can be realized only with highly reliable wireless transmission. As the  \SI{5.9}{\giga\hertz} frequency band used for V2V suffers from high congestion, in this paper, we consider the use of the terrestrial TV frequencies for intra-platoon communications. In order to be able to evaluate the potential of the new bands fully, propagation models for V2V communications at such frequencies are needed. Therefore, this paper reports new V2V propagation measurements and their modeling results. Particularly, we propose a Double Slope Double Shadowing model as the most accurate one, based on a comparison of various models using the Bayesian Information Criteria. We also investigate the space-time autocorrelation properties of the shadowing, which turned out to be dependent on the speed of vehicles. The proposed path loss and shadowing model differs from the ones proposed for the \SI{5.9}{\giga\hertz} band. Mostly, in favor of the TV band, as shown by, e.g., no statistically significant impact of a blocking car.
\end{abstract}

\begin{IEEEkeywords}
path loss, shadowing, modeling, propagation, TVWS.
\end{IEEEkeywords}

%
\IEEEpeerreviewmaketitle

\section{Introduction}
%
%
%
%
\IEEEPARstart{W}{ith} the increase in road traffic nowadays, provisioning of transport safety and reliability is of utmost importance. Various inventions have been introduced in the automotive industry in the past years, just to mention, e.g., Anti-lock Braking Systems (ABS), Anti Slip Regulation (ASR), Electronic Stability Program (ESP), lane assist systems, pre-collision assist or Adaptive Cruise Control (ACC) systems. 
Further advancements in the safety and reliability of cars led to the formulation of new specific use cases, among which platooning is particularly interesting, offering potentially significant gains in traffic efficiency and safety. Vehicle platooning is a coordinated movement of a group of closely following vehicles that can be driver-controlled or autonomous, forming a convoy led by a platoon leader. Among the envisaged gains of platooning, one can identify the increase in road capacity, reduction of fuel consumption, and improvement of driver safety and comfort. Studies have shown that platooning may double the utilization of road capacity due to a reduction of inter-vehicle spacing \cite{LPT+16}. Moreover, the findings of the
SARTRE project shows that platooning provides up to 15\% fuel savings for trucks traveling behind the platoon
leader \cite{EChen_2012}. That would yield significant benefits for transport companies. Furthermore, the fuel savings translate to a substantial reduction of the carbon footprint on the environment. According to findings of the Energy ITS project \cite{Tsugawa_2011}, when the market penetration of truck platooning increases to 40\% of trucks, the CO2 emissions along a highway can be reduced by 2.1\% if the gap
between trucks is \SI{10}{\metre}, and by 4.8\% if the gap is reduced to \SI{4}{\metre}.

As stated above, the main factor determining the benefits of platooning is the inter-vehicle gap that influences the air drag and the road capacity. In order to keep the distances between vehicles minimal, advanced driver assistance systems or autonomous controllers, such as the Cooperative Adaptive Cruise Control (CACC), need to be considered for the platoon cars that require ultra-reliable low-latency (URLLC) communications. According to the 3GPP specification for 5G, the required reliability level of wireless communications for platooning may reach even 99.99\%, with the latency between \SIrange{10}{25}{\milli\second} \cite{3gpp2018b}. Two wireless communication standards are currently considered for vehicle-to-everything (V2X), namely the Dedicated Short-Range Communications
(DSRC), relying on the IEEE 802.11p specification, and the Cellular V2X (C-V2X), being an extension of LTE standard (LTE-V2X) or 5G New Radio (NR-V2X). Both are capable of providing sufficient communications quality; however, like most wireless systems, their transmission reliability is prone to the effect of wireless channel congestion and limited range. According to the studies presented in \cite{VUKAD2018}, when using the 5.9 GHz frequency band dedicated for V2X communications, such an assumption of sufficient transmission quality may not be valid in dense traffic scenarios, especially in the case of platooning, where high messaging rate and reliability is required. Therefore, alternative transmission opportunities have to be sought.

Among the widely considered concepts for solving the problem of congested wireless channels, the utilization of the unused (or rarely used) licensed frequency resources, known as the white spaces (e.g., the television frequency band white spaces - TVWS), in the form of Dynamic Spectrum Access (DSA) has been proposed in the past, e.g., for broadband Internet access \cite{Kliks_13, JPR2015}. According to \cite{SYB2018, Sroka2020a}, a similar approach can be applied to V2X communications to increase the reliability of intra-platoon communications, with the TVWS utilized dynamically. One of the main advantages of the TVWS is that the location of the Digital Terrestrial Television (DTT) transmitters, as well as their configuration, is relatively stable on a long time scale and can be easily predicted based on location. Therefore, with the dynamic allocation of the available TVWS frequency bands, the needed bandwidth might be provided to platoons for exclusive or shared use, resulting in separation from other V2X services, thus improving communications security and robustness. Furthermore, the use of  TVWS frequencies may result in an increase in transmission range, compared to the \SI{5.9}{\giga\hertz} band, as the propagation effects, such as the shadowing due to the obstacles (e.g., vehicles) or the signal attenuation due to the distance, are less severe when using lower frequencies. In our work, we concentrate on frequencies between \SI{470}{\mega\hertz} and \SI{862}{\mega\hertz}, typically devoted to TV transmission, and in particular, we perform our experimentation study at a center frequency set to \SI{725}{\mega\hertz}. The applicability of the proposed method has to be adjusted to the regional regulatory rules, e.g., those addressing the ITU recommendations on the digital switch-off done in last years in various countries all over the world. Transmission in the TVWS is characterized by significantly lower attenuation than in the \SI{5.9}{\giga\hertz} band, which can be even up to \SI{20}{\deci\bel} lower, e.g., in the case of Line-Of-Sight (LOS) transmission. On the other hand, one should keep in mind that the considered frequency range is licensed; thus, the primary DTT system needs to be protected. Moreover, channel availability may change depending on the location. Although numerous measurement campaigns have shown the prospective availability of TV bands for secondary transmission, there may be areas where the number of occupied TV channels is high \cite{Kliks_13,Hessar2015}. However, as the platoons will usually move on the motorways and expressways that are located far from potential DTT receivers, and due to their high mobility (only short-term interference is expected), the protection requirement can be fulfilled; however, it still needs to be taken into account. Hence, the use of a dedicated database subsystem storing information describing the spectrum awareness in certain locations, also known as Radio Environment Map (REM), can help with the identification of the prospective bands and adaptation of the transmission parameters in order to protect the primary system \cite{Kliks_13,JPR2015, Sroka2020a}. Such a database-oriented system will also be responsible for storing the information about the areas, where only communications with classic DSRC may be realized. 

One of the main problems when evaluating the usefulness of TVWS for V2X is the lack of reliable propagation models between two vehicles for lower frequencies (VHF and UHF bands). The existing models for V2V usually assume communication in the 5.9 GHz band. On the other hand, the models applicable for the considered frequency range usually assume that either the transmitter or the receiver operates with much greater antenna height (e.g., on a mast), which makes it unreliable when considering V2V propagation. Therefore, the goal of this paper is to facilitate the potential deployment of V2V communications in the TVWS band by proposing a large-scale fading model for this use case, i.e., the path loss model that considers the presence of shadowing \footnote{The large-scale fading considers both effects of path loss and shadowing. However, in vast papers in the case of the real-data measurements, the term \textit{path loss} is used to describe signal power attenuation between transmitter and receiver. This is the same as large-scale fading in this context. In this paper, we focus on the large-scale fading model; however, in common sense, our paper addresses the path loss modeling where the shadowing effect is inherently included}. Thus, the TVWS channel model development is of the highest necessity as long as the accurate and reliable results are needed to investigate this frequency band (simulation results for new V2X solutions, device development, database communication, etc.).

Extensive measurements were carried out using a channel sounder built at Poznan University of Technology. The whole setup, including the software and hardware used, is presented. The measurements were conducted in normal traffic conditions, with varying speed, the distance between cars, and the number of obstacles. The collected samples are tested against four hypothetical large-scale fading models. Maximum Likelihood estimator with samples censoring \cite{Gustafson_ML_estimation_censoring_2015} is used for parameters estimation. Next, the Bayesian Information Criteria (BIC) is used to find the most suitable model. Moreover, shadowing samples autocorrelation is studied from distance and time perspectives, discussing the influence of other factors, e.g., vehicles' speed. This is an important aspect for enabling modeling and simulations of V2V communications. 

The main novel aspects and contributions of this paper can be summarized as follows:
\begin{itemize}
    \item A double slope double lognormal shadowing path loss model for inter-vehicle communications in the terrestrial TV band is proposed, based on extensive measurements. Four different models have been considered, with the chosen one providing the highest accuracy according to Bayesian Information Criteria and simplicity of calculation.
    \item  The results confirm that TVWS are a good choice for intra-platoon communications considering the lack of statistically significant signal blockage by the cars or propagation conditions not changing significantly in time. 
    \item The effect of shadowing autocorrelation is analyzed, taking into account the mobility parameters of the transmitter and the receiver. The impact of cars' velocities is captured with the proposed time-based correlation model, which turned out to be more accurate than the distance-based model.  
    \item The main aim of the developed propagation model is to provide a simple and accurate solution, applicable for different traffic and environmental conditions, which can be used as a tool for providing estimates of channel impact for self-organizing systems, such as the REM-aided platooning proposed, e.g., in \cite{SYB2018,Sroka2020a}.
\end{itemize}

The paper is organized as follows. Section II provides a state-of-the-art overview of the available V2V channel models. In Section III, a detailed description of the measurement experiment setup is given, with an explanation of the post-processing applied to the obtained results. Section IV outlines the details of the carried out measurement campaign, while the development of the proposed path loss models is described in Section V. Finally, Section VI provides the summary of validation of the proposed models with additional experiments, and Section VII concludes the paper.

\section{State of Art in V2V Wireless Channel Attenuation Modeling}
The wireless propagation characteristics between vehicles have been well investigated in the case of transmission at a \SI{5.9}{\giga\hertz} carrier. This is a result of DSRC standard popularity and band allocated at \SI{5.9}{\giga\hertz} band by ITU for this application. While the propagation is affected by different phenomena, the developed models usually capture the average channel attenuation separately as a path loss component dependent, e.g., on the transmitter-receiver distance and the local variations (e.g., due to the local situation of the transmitter, receiver, and the surrounding elements) in the form of shadowing, usually represented as a zero-mean random variable. The shadowing variable can be correlated in time or distance, so that dependence on the speed of vehicles is included. A problem not covered in this manuscript is the modeling of fast-fading effects for V2V channels or time-evolving clusters of paths \cite{cai_2018}.

First solution to mention is \cite{Cheng_5_9GHz_Channel_model}. The authors made single carrier measurements in a suburban environment in Pittsburgh, PA. Two measurements were carried out: in the first data set wireless channel is observed over \SI{1.5}{\second} (per location), and in the second data set, the wireless channel is observed over \SI{0.2}{\second}. The path loss is modeled using single and double slope models (separate models for distances shorter and longer than \SI{100}{\metre}) supplemented with normally distributed shadowing. For a single slope model, path loss exponent equals \num{2.75} or \num{2.32}, and shadowing standard deviation equals \SI{5.5}{\deci\bel} or \SI{7.1}{\deci\bel}. The double slope model uses a path loss exponent of about 2 for the short link (below \SI{100}{\metre}) and around 3.9 for longer links. In addition, the authors analyzed small-scale fading (showing Rician fading for short links and Rayleigh fading for long links), dependence between vehicles' speed, Doppler spread, and coherence time. 

In \cite{abbas2015measurement} wideband V2V channel sounding measurements at the carrier frequency of \SI{5.6}{\giga\hertz} are reported for urban and highway environments in Sweden. Most importantly, a video recording of the path between the TX and RX car was captured in parallel to RF and GPS measurements. This has been done in order to tag the RF measurements with the type of the link, i.e., Line of Sight (LOS), Obstructed LOS (OLOS), when the direct visibility between cars is obstructed at least partially by another vehicle, and Non-LOS (NLOS) when there is a building blocking LOS between TX and RX. The measured power was averaged locally over all measurements obtained within 15 wavelengths (of measurement car route) to remove fast fading effects. For each environment, a separate double slope path loss model was proposed with a breakpoint distance (separating both slopes) of \SI{104}{\metre}. It is observed that OLOS propagation results in around \SI{9}{\deci\bel} stronger attenuation than obtained in the LOS environment. In addition, the OLOS can be characterized by higher shadowing standard deviation than LOS propagation. Finally, the authors derive long-term fading (shadowing) decorrelation distance based on measurements. It equals about \SIrange{20}{30}{\metre} for highway scenario (depending on link type) and around \SI{4}{\metre} for the urban environment. 

Among the advanced tools making use of the stochastic geometry, a commonly used V2V channel propagation model for \SI{5.9}{\giga\hertz} is the Geometry-based Efficient propagation model for V2V communication (GEMV$^2$) proposed in \cite{Boban_TVT_2014}, \cite{Chebil2021}. It requires modeling the shape of each building and car in the considered scenario. Each link is classified as LOS, NLOSv (Non-LOS because of vehicles), and NLOSb (Non-LOS because of buildings). While the LOS link uses a two-ray ground reflection model, NLOSv considers vehicles being obstacles and diffraction on them. NLOSb uses a log-distance path loss model with a distance exponent of \num{2.9}. The proposed model implementation is freely available on the web. However, for the sake of simplicity, throughout the rest of the paper, the LOS, NLOS, and OLOS will be used to differentiate between types of radio channels.

The important radio propagation measurements from the perspective of intra-platoon communications at \SI{5.9}{\giga\hertz} are reported in \cite{Nilsson_2017_V2V_model}. The paper analyzes path losses between 4 cars constituting a convoy on a highway. The two-ray ground reflection model is used for LOS links and a single slope model for OLOS links. An important aspect is the analysis of log-normal shadowing: how it is correlated in space for a single car and how it is correlated between cars.

Also, other papers consider the aspects of channel modeling in the context of platoon-type communication. E.g., \cite{yang2021}, \cite{Yang2021TVT} present channel modeling for three complicated scenarios: viaduct, tunnel, and cutting in the \SI{5.9}{\giga\hertz} or \cite{Eckhardt2021} presents and discusses measurements in the low terahertz band.



All the above-mentioned models consider frequencies at least 10x higher than typical TVWS frequencies. This significantly influences basic signal propagation properties, e.g., the ability to propagate through objects, diffraction, or reflection characteristics. 
However, there are hardly any works proposing models for V2V communications in TVWS. One of these is \cite{Kitamura_RayTracing_V2V_TVWS_2018}, which uses Ray-Tracing to model signal power and delay distribution at a junction. Unfortunately, the authors do not propose any path loss or shadowing model. In \cite{Arteaga_V2V_TVWS_Hata_2018} authors compare reliability in message distribution between vehicles at \SI{5.9}{\giga\hertz} and in TVWS. They propose to use the Hata model for V2V path loss modeling. Unfortunately, while the Hata model covers the TVWS band, it is not validated for both transmitter and receiver being located at low height with a distance ranging from a few meters to a few hundred meters. Due to this fact, results obtained with the use of this model for the V2V communication where the antenna heights are in the range of \SIrange{1}{3}{\metre} (cars/trucks) will make the results questionable.

This motivates us to perform extensive path loss and shadowing analysis for TVWS in the V2V communications context, which bases on the measurements conducted in real traffic conditions. 

\section{Measurement Setup and Signals Post-processing}
\subsection{Used Equipment and Connections}

\begin{figure}[!hbt]
\centering
\includegraphics[width=3.4in]{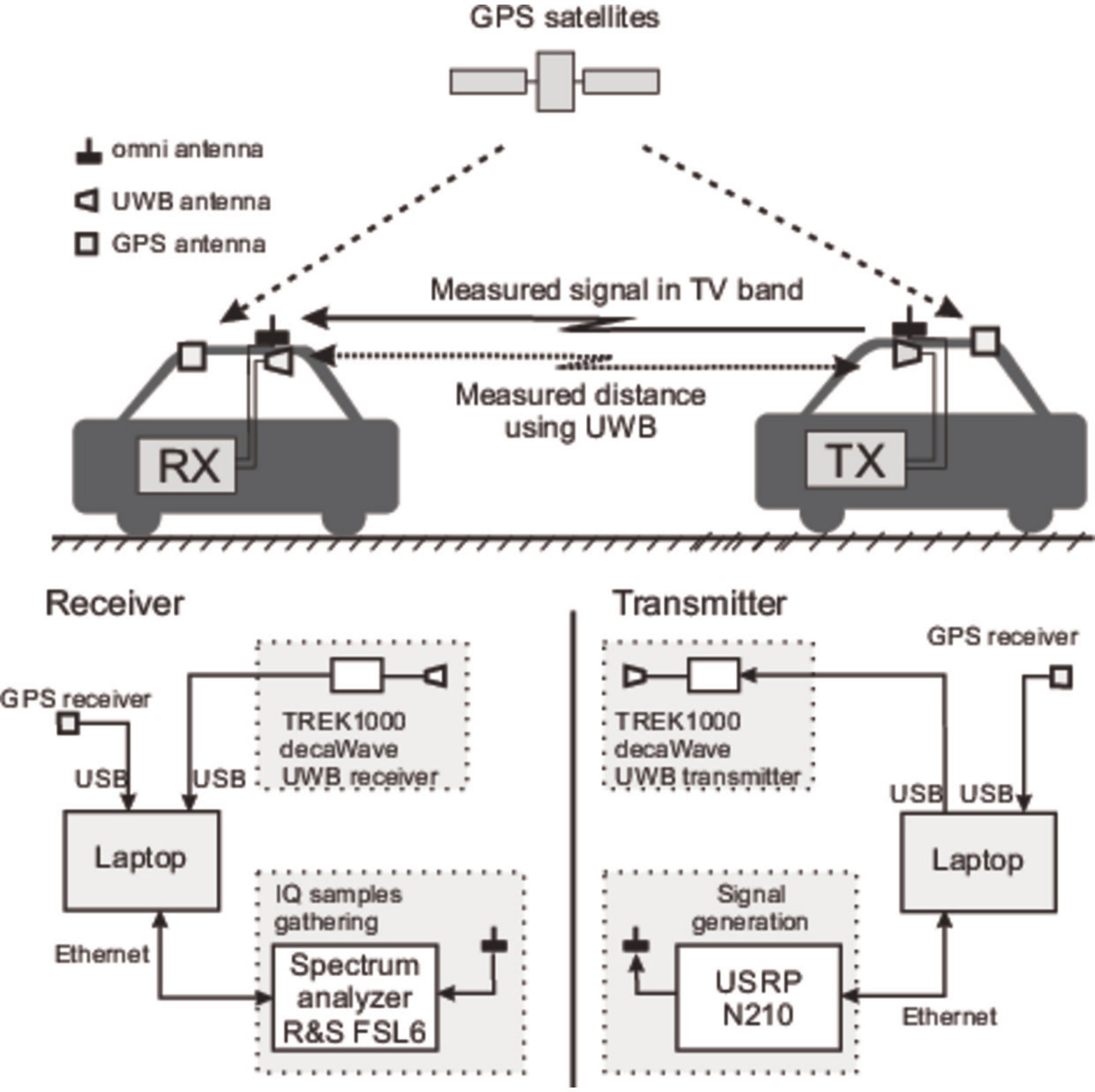}
\caption{Measurement setup}
\label{fig_measurement_setup1}
\end{figure}

The channel measurements have been carried out using a setup composed of two cars with radio and localization equipment as shown in Fig. \ref{fig_measurement_setup1}. 

One of the cars, denoted from now as the TX car, transmits a single multicarrier symbol in a loop. The complex symbol is transmitted with a sample rate of \num{8.33}~MSps. It has 200 out of 256 subcarriers modulated, meaning that the useful bandwidth sounded by the comb of sinusoids equals \SI{6.5}{\mega\hertz}. The symbols modulating each subcarrier are optimized in phase according to \cite{Channel_Sounding_Neul} so that the Peak to Average Power Ratio (PAPR) is minimized, obtaining \SI{0.73}{\deci\bel}. The transmission is carried out using a laptop with GNU Radio installed. The IQ samples are transferred to a Software Defined Radio Platform: USRP N210 with a WBX frontend. Additionally, the GPS-disciplined oscillator is installed in USRP in order to provide high accuracy and stability of carrier frequency. The utilized carrier frequency is set to \SI{725}{\mega\hertz}. The output RF signal is further amplified using Mini Circuits ZRL3500+ and fed to magnetic rooftop-mounted antenna PCTEL LPBMLPVMB/LTE of gain \SI{3}{\deci\bel i}. The Effective Isotropic Radiated Power (EIRP) has a mean\footnote{ While the transmitted sequence has varying in time samples power, the mean over time is used for EIRP calculation.}
of \SI{22.2}{\deci\bel m}.
The TX car is additionally equipped with a GPS receiver that logs NMEA data once every second. The third element is the Decawave TREK1000 Ultra Wideband (UWB) device that is used for precise distance estimation between both cars.   

The second car, denoted as the RX car, runs Rohde\&Schwarz FSL spectrum analyzer connected to an omnidirectional antenna PCTEL LPBMLPVMB/LTE of gain \SI{3}{\deci\bel i} that is mounted on the car's roof. The spectrum analyzer is set to IQ trace mode, allowing to fetch IQ samples by the laptop connected via Ethernet cable. The configuration is adjusted to maximize sensitivity (enlarging the channel-sounding range) while not observing signal overloading spectrum analyzer input. It was obtained by turning on the preamplifier and setting the internal attenuator to \SI{15}{\deci\bel}. The laptop runs Matlab, which manages the spectrum analyzer measurement together with data obtained from the GPS localization receiver via a dedicated script. The GPS data are obtained using the NMEA protocol once every second. In addition, the TREK1000 device runs and logs the distance to the device installed on the TX car. While the time-of-flight is out of our interest, the TX and RX operate in an asynchronous way, i.e., there is no synchronized triggering of the TX and RX device.

{\PK The antennas installation points have a significant impact on the observed results. We decided to mount the TX and RX antennas on vehicles roofs. This can be a real-life case as it simplifies installation thanks to typical vicinity of the Frequency Modulation (FM)/Digital Audio Broadcasting (DAB) radio receiving antennas. At the same time this should minimize the attenuation introduced by intermediate cars as reported for \SI{5.9}{\giga \hertz} band \cite{Boban_TVT_2014}, improving reliability of V2V communications. If a multi-hop transmission is considered, especially in mmW, the most suitable antenna position can change e.g., to bumper level~\cite{Kryszkiewicz_Softcom2021}.} 
As known, a metallic roof has a meaningful impact on antenna characteristics \cite{mocker_2015}. However, in the case of real usage, the same phenomena will be observed. Therefore, we decided not to distinguish between the roof-mounted antenna and the metallic roof (due to the antenna vicinity could also be considered a part of the antenna). The unavoidable effects of a roof on propagation will be reflected by the measured channel characteristics.

\begin{figure}[!t]
\centering
\includegraphics[width=3.4in]{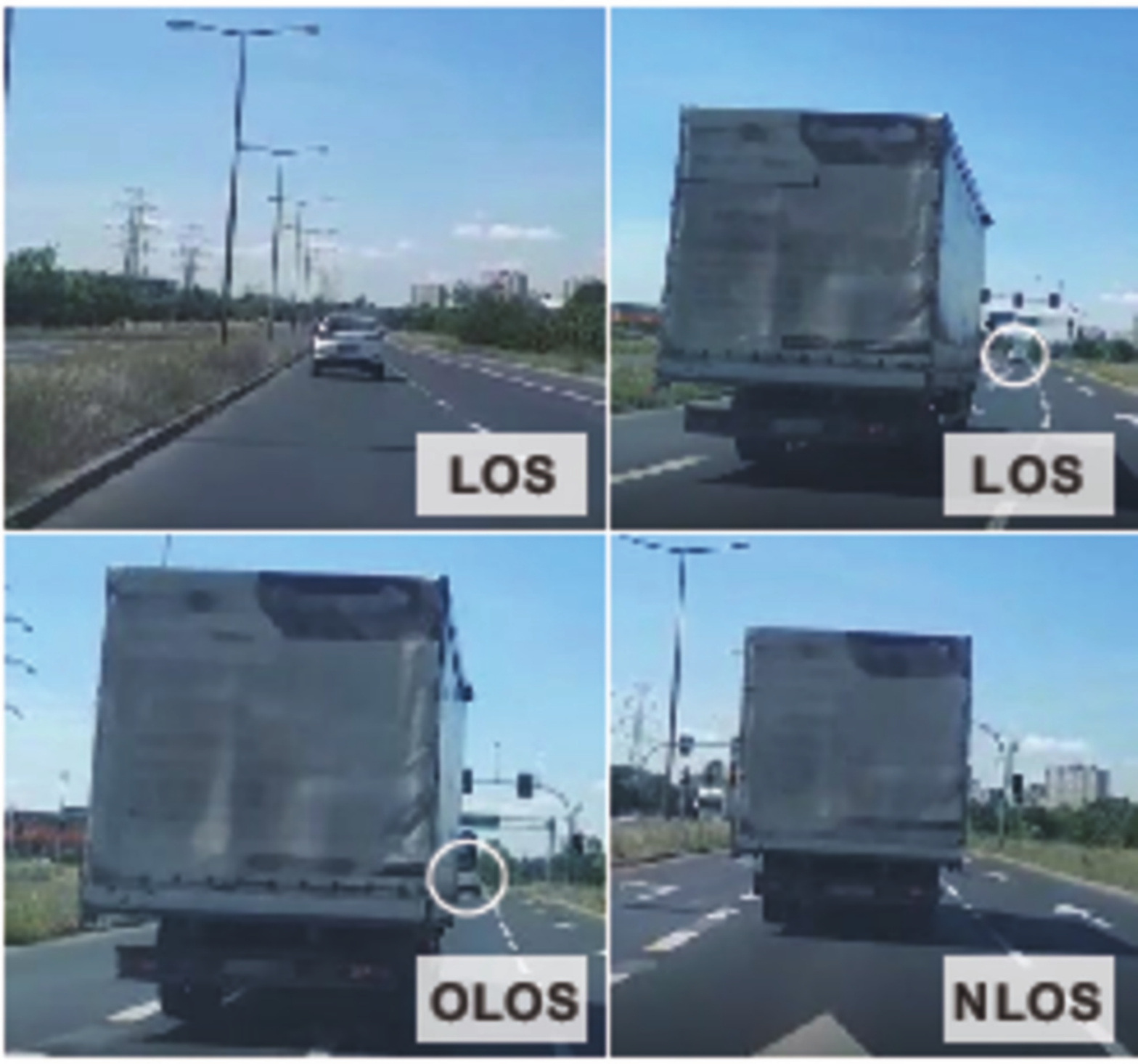}
\caption{Illustration of various path loss scenarios: line of sight (LOS), obstructed LOS (OLOS), and Non-LOS (NLOS); }
\label{fig_losolosnlos}
\end{figure}

\begin{figure}[!t]
\centering
\includegraphics[width=3.4in]{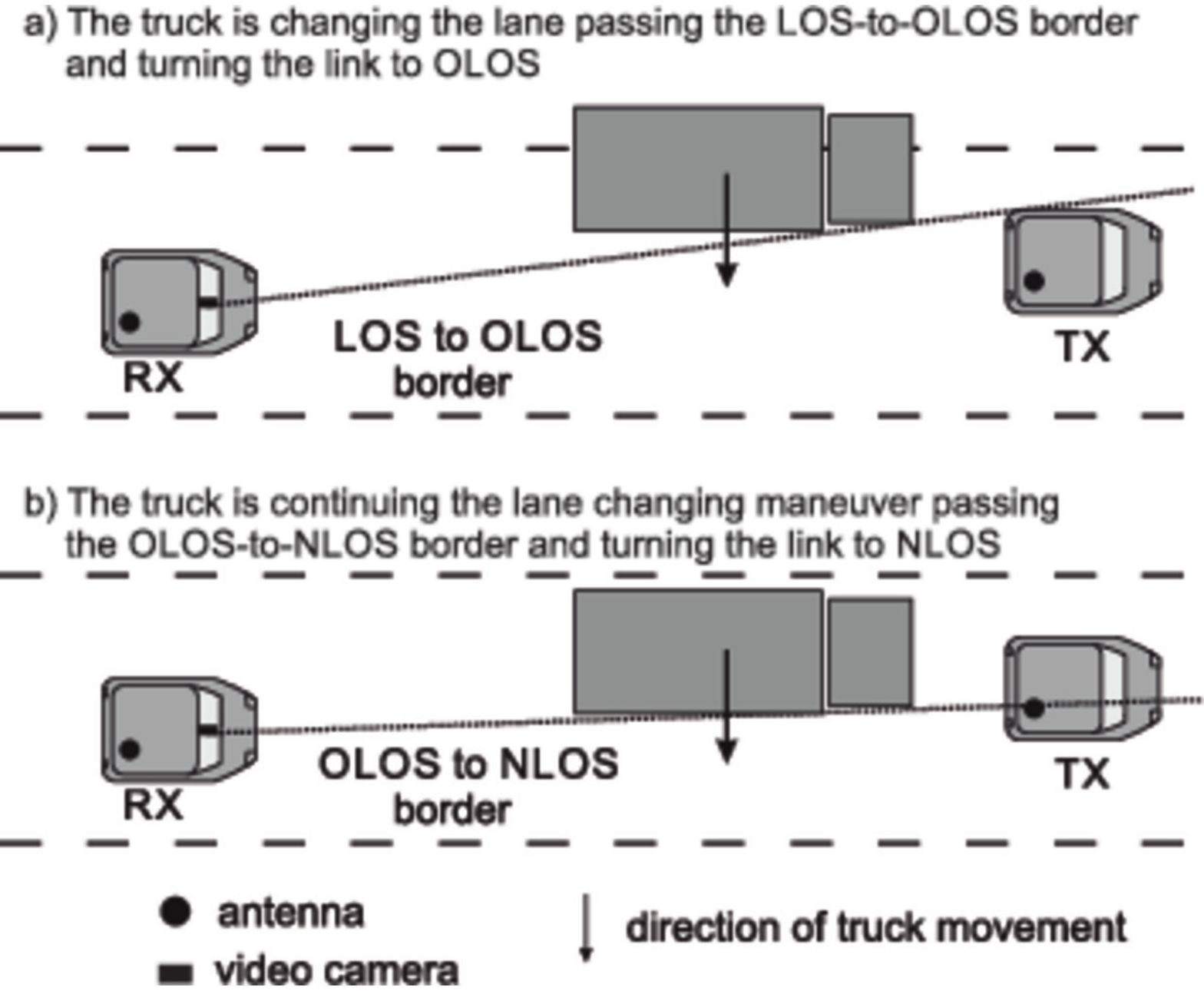}
\caption{Detailed illustration of the transition between LOS, OLOS and NLOS; The truck is changing the lane entering the space between the transmitter and the receiver}
\label{fig_losolos_transition}
\end{figure}

In order to support the post-processing of all measured data, all routes have been recorded using a dedicated camera mounted just behind the front window of the RX car. Then, these recordings have been analyzed second by second to distinguish between the various link conditions. Following, e.g., \cite{abbas2015measurement}, we have identified three path loss cases (see Fig.~\ref{fig_losolosnlos}):
\begin{itemize}
    \item  Line-Of-Sight (LOS), where there is no obstacle between the transmitting and receiving cars, and the direct line of signal propagation can be identified. Though, this is not equivalent to free-space propagation;
    \item Obstructed Line-Of-Sight (OLOS), where the link between the transmitting and receiving cars is partially obstructed by another car; 
    \item Non-Line-Of-Sight (NLOS), where there is no LOS between the transmitting and receiving cars.
\end{itemize}
Theoretically, the just way to define the impact of the blocking car on the transmission is to incorporate the first Fresnel zone - i.e., if the blocking car is within the range of 0 to 0.8 of the first Fresnel zone radius, then it will be treated as obstructive. In practice, however, in the normal traffic conditions, where the measurements have been carried out, it is impossible to derive precisely the location of the blocking car in relation to the transmitting and receiving one. In consequence, we have to approximate the impact of the blocking car by applying visual post-processing of the recorded videos. Thus, when the whole transmitting car can be recognized in the videos, the link was treated as LOS. Otherwise, we treat the link as blocked, and further subdivide it into OLOS and NLOS. 
We start treating the link between the transmitting and receiving cars as OLOS when the blocking car starts overlapping visually with the preceding car (regardless of the direction from which the blocking car has arrived and regardless of the relative location of the transmitter and receiver, i.e., if they were on the same lane or the neighboring lanes). We treated the channel as OLOS until most of the transmitting car was hidden behind the blocking car. In particular, the transition between OLOS and NLOS was done, when around 80 - 90~\% of the transmitting car was visually overlapped by the blocking car. This idea is illustrated in Fig.~\ref{fig_losolos_transition}, where the example of a typical maneuver is illustrated. In particular, the truck is changing its lane (in the figure from top to bottom), entering the space between the transmitter and the receiver, obstructing, in consequence, the radio link between them and causing the change of its type. In the top subfigure, the first phase of the maneuver is shown, where the truck is passing the LOS-to-OLOS border. Consequently, the link between the transmitter and receiver is no longer of LOS type and will be treated as OLOS. The OLOS link will exist as long as the blocking truck passes the OLOS-to-NLOS border, turning the link into a non-line-of-sight. This situation is illustrated in the bottom subfigure. One can notice that the visual distinguishing between the LOS, OLOS, and NLOS scenarios may be biased with some errors, e.g., when the cars are so small in the video that it is hard to separate the LOS-to-OLOS and OLOS-to-NLOS borders precisely. In such a case, we apply the most restrictive approach, i.e., we treat the link as LOS or OLOS only when we are visually sure that such kind of link exists.


\subsection{Synchronization and Distance Estimate}
An essential aspect is the synchronization of all three measurements, i.e., RF signal, GPS localization, and UWB-based distance. Assuming that in the worst case 2 cars travel from the opposite directions at 100 kmph speed, their distance change by \SI{55}{\metre} every \SI{1}{\second}. As the distance is expected to have the most significant influence on the path loss, the RF channel measurements should be aligned in time with the distance between cars, obtained within at least the same decimal part of a second. First, the laptop recording all data is synchronized using a time server. This results in each RF and UWB measurement being tagged with the system time of accuracy within a single ms. On the other hand, the GPS location information is provided once a second, at the beginning of each second, together with this timestamp. As this happens at both link ends (according to NMEA specifications), Vincenty's algorithm is used to calculate the distance between two coordinates obtained at exactly the same time instants. Next, the distances obtained for consecutive seconds are interpolated using linear interpolation to get the inter-vehicle distance at the time of RF signal reception. 

The UWB distance measurements are performed with the use of TREK1000 from Decawave. The TREK1000 is a wireless transceiver operating according to {IEEE 802.15.4-2011} standard that allows for range measurement with high accuracy. This device can operate in \SI{3.993}{\giga\hertz} or \SI{6.486}{\giga\hertz} bands with the data rate of 110~kbps or 6.8~Mbps with the expected accuracy of  \SIrange{\pm 10}{30}{\cm} (depending on the scenario). The UWB measurement setup consists of one device in the TX vehicle (only powered via USB) and the second module in the RX car that was connected to the laptop for distance logging. UWB transceivers, operating inside each car, were connected with the rooftop-mounted TX/RX antennas with the ca.~\SI{1.8}{\metre} long RF cables. These cables caused the reported distance to be increased by \SI{5}{\metre} in comparison to reality~\cite{Kryszkiewicz2020}. Therefore, the value of \SI{5}{\metre} was subtracted from the distance reported by TREK1000 to provide an unbiased, UWB-based inter-car distance. The UWB receivers report distance about \num{3.5} times per second, while the regular GPS receiver reports distance once every second.

\begin{figure}[hbt]
    \centering
    \includegraphics[trim=0.5cm 0.5cm 0.5cm 0.7cm,width=0.5\textwidth]{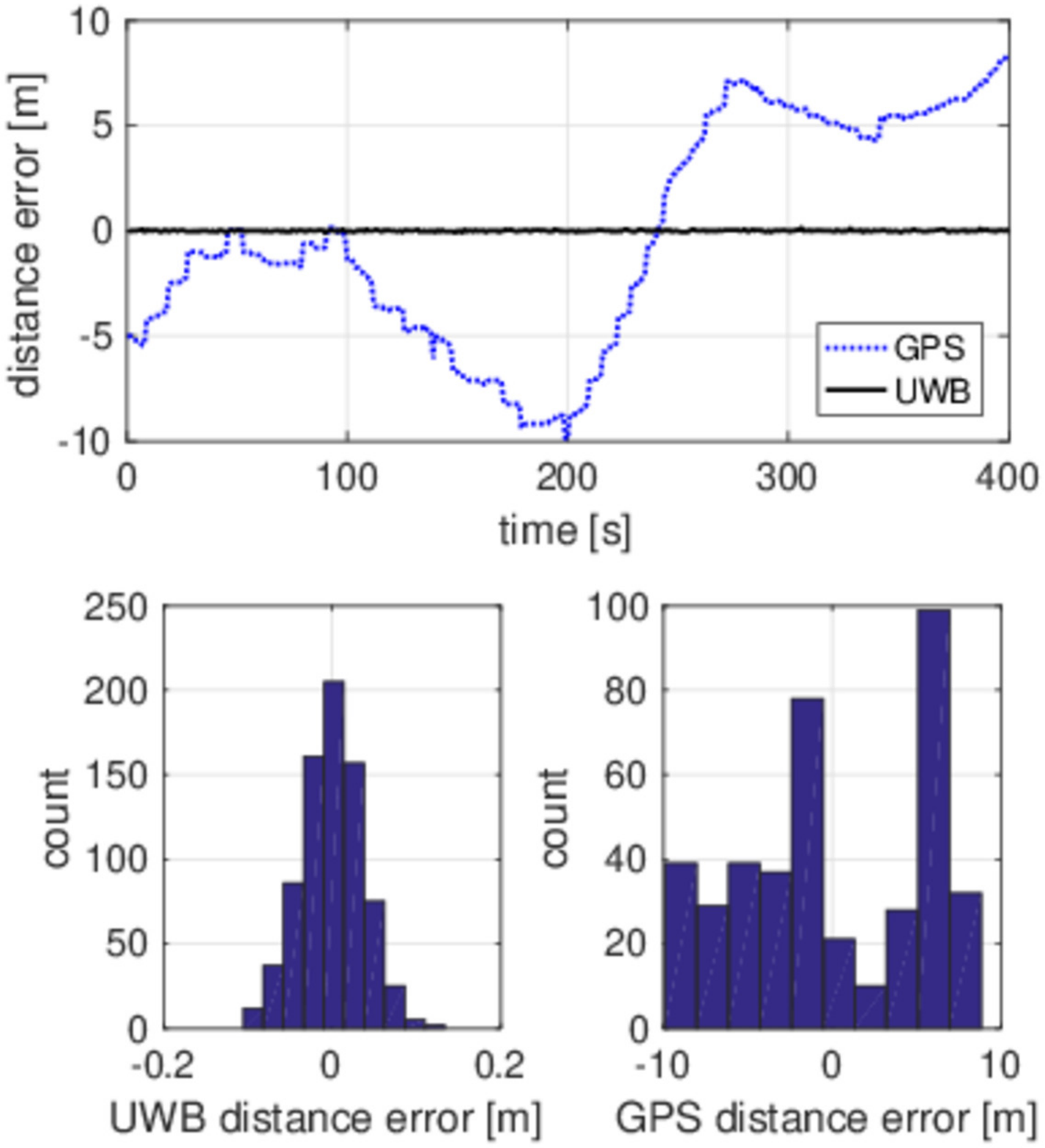}
    \caption{Fluctuations of the GPS and UWB distance error around the mean value}
    \label{fig:stability}
\end{figure}

The need for UWB-based distance measurement to complement GPS-based measurement results from the characteristics of these two technologies and the requirements of V2V channel modeling. The GPS allows for distance estimation between any two locations on the globe. However, these measurements are characterized by a relatively low accuracy. As reported in \cite{Kryszkiewicz2020} a few meters standard deviation can be expected. Even application of more sophisticated devices, e.g., u-blox, slightly improve the measurement accuracy. At the same time, the GPS-based distance error is significantly correlated in time. This prevents consecutive GPS-based measurements from being averaged to remove the error. An example of instantaneous distance measurement error in time is presented in Fig.~\ref{fig:stability}. It is obtained for both TX and RX vehicles parked at a fixed distance of \SI{44}{\metre}. It is visible that the maximal absolute distance error reaches nearly \SI{10}{\metre} and maintains this value over several seconds.   
During the same time, UWB-based distance estimates have been observed. The standard deviation of the error equals \SI{0.036}{\metre} (while ca.~\SI{5.4}{\metre} for GPS-based distance estimates). This is reflected by the plot of error change in time and its histogram in Fig.~\ref{fig:stability}. However, contrary to GPS-based measurements, measurements carried out using the UWB technique are limited to the range of several hundred meters (effectively several dozen).
Fortunately, the V2V channel modeling requires more accurate distance estimates for the relatively low distance between vehicles. This comes from the typically logarithmic influence of distance on the modeled path loss as justified in \cite{Kryszkiewicz2020}.



Therefore, the accurate UWB-based distance estimates, available only for relatively short links, are used to remove the instantaneous distance estimation error obtained by GPS-based measurement using an algorithm proposed in \cite{Kryszkiewicz2020}. If both UWB and GPS-based measurements are available for a given time instance, their results are combined using the maximal ratio combining method (with weighting factors dependent on the standard deviation of UWB and GPS measurements). The algorithm takes advantage of the relatively long decorrelation time of the GPS-based measurements, i.e., a several seconds long period over which the GPS-based measurement error is stable as visible in Fig.~\ref{fig:stability}. The UWB-based distance correction is assumed to be valid approximately \SI{40}{\s} before and after each UWB measurement, with a gradually decreasing influence on the final estimate. 
If the accurate UWB-based measurements are not available over a longer period, the correcting factor cannot be estimated, and the GPS-only measurement is used. While the unavailability of the UWB-based measurement is typically caused by the large TX-RX distance, this is acceptable for the path loss modeling. This is caused by a logarithmic influence of distance on path loss. This issue is discussed in detail in \cite{Kryszkiewicz2020}. Another problem with UWB-based distance measurement is the degradation of its performance in the OLOS/NLOS channel. However, the expected accuracy is still better than in pure GPS-based distance measurements~\cite{Wu_UWB_NLOS_2007,Yang_UWB_NLOS_2018}. Additionally, the OLOS/NLOS channel is typically reported for medium or long TX-RX distance. As mentioned above, for these distances, the UWB-based measurement is not required to keep a high accuracy of path loss modeling.



\subsection{Path Loss Estimation and Post-processing}
\label{sec_pathloss_est}
The spectrum analyzer saves 1024 consecutive IQ samples in each run. Considering a sampling frequency of 8.33~MSps this is equivalent to \SI{0.123}{\ms} of continuous channel sounding. These samples are time and frequency synchronized to the transmitted sequence utilizing its autocorrelation properties according to \cite{Channel_Sounding_Neul}, and Signal to Noise power Ratio (SNR) is estimated. The utilized waveform combined with the detection method achieves a sensitivity of \SI{-9.9}{\dB} SNR (for the probability of false alarm equal 1\%) that can be converted to \SI{-85.4}{\dB m} power level considering spectrum analyzer noise power. While the TX uses an EIRP of \SI{22.2}{\dB m} and the receiver uses an antenna of \SI{3}{\dB} gain, the maximal path loss that can be estimated without significant noise-based error is \SI{110.6}{\dB}. This is used as a censoring level required for the Maximal Likelihood path loss modeling according to the method described in \cite{Gustafson_ML_estimation_censoring_2015}. 

Calculating mean incoming signal power (by averaging squared magnitudes of incoming IQ samples) and utilizing the estimated SNR allows for computing the wanted signal power. Because of the time required by the spectrum analyzer for internal signal processing and data transfer to the connected computer, the new channel-sounding vector is obtained approximately every \SI{16.5}{\ms}. Assuming the measuring car moves at most \SI[per-mode=fraction]{100}{\km\per\hour}, a new measurement is obtained every \SI{0.46}{\m}. This is similar to a single wavelength ($\lambda=0.41 m$ for a carrier frequency of \SI{725}{\MHz}). In order to remove the fast fading effect from the measurements, two types of local averaging are used, i.e., frequency-domain averaging thanks to the utilization of wideband measurements and time domain averaging of 10 consecutive power measurements (in linear scale) as in \cite{Nilsson_2017_V2V_model}. This is in line with a typical rule of thumb that fast fading is a phenomenon that can be removed by averaging measurements obtained over a few or a few tens of wavelengths. In \cite{Lee_fading_removal_1985}, following the analytical derivations for the Rayleigh channel, a 40-wavelength averaging is suggested. However, the measurements reported in this paper are carried with a fixed time period but variable speed, changing the distance between measurements and making the exact 40 wavelengths averaging infeasible. Moreover, the Rayleigh fading assumption may not be valid for many measurements, e.g., if there is a LOS component. Still, the averaging in the frequency domain combined with 10 times averaging in the time domain should suffice to remove fast fading effects.
As a result, the path loss estimate is tagged with the time and distance between both cars obtained from GPS and UWB data, as explained above.

\section{Measurement Campaign}
\label{sec_meas_campaign}
In order to investigate the large-scale fading model (i.e., the model that covers both path loss and shadowing) for TV bands in the context of V2V communications (mainly, but not limited to platoons), extensive measurement campaigns have been carried out on various days, day-time, and various traffic conditions. In all cases, two cars have been involved in the experiment, as illustrated in Fig.~\ref{fig_measurement_setup1}, mainly Hyundai Tucson (acting as the transmitter) and Peugeot 5008 (acting as the receiver). Both the transmit and receive antennas were mounted on the assumed height of \SI{1.75}{\m} above ground, i.e., $h_{TX}=h_{RX}=1.75$m. Moreover, at approximately the same height (down to some centimeters), the UWB antennas have also been mounted. The video camera was installed behind the windshield, in the car's center. As for the data storage, we have measured the IQ samples directly and saved these values together with a detailed location from the GPS receiver. There were in total 496 files, each containing 1000 vectors of 1024 samples.

\begin{figure}[!htb]
\centering
\includegraphics[width=3.4in]{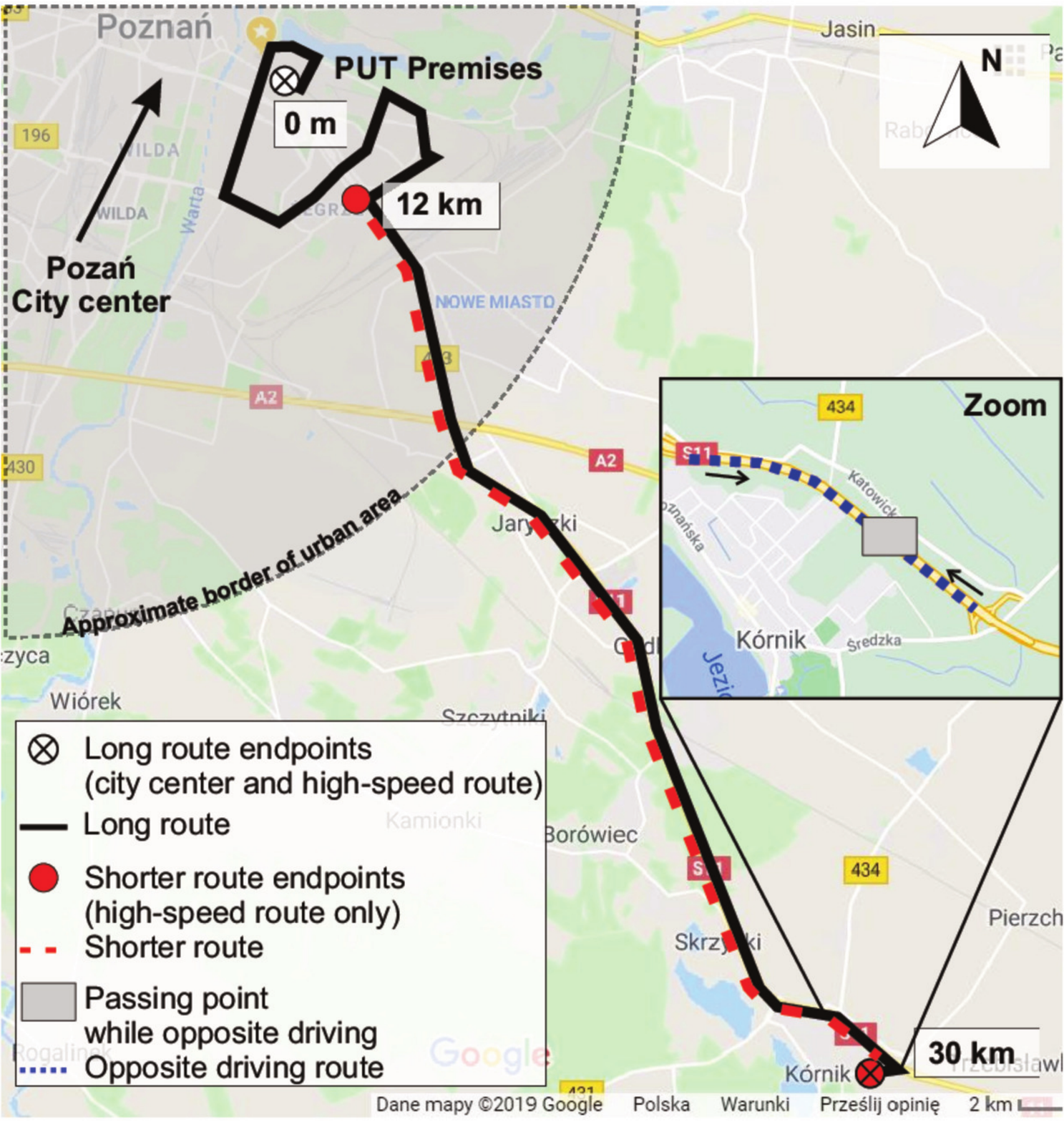}
\caption{Measurement scenarios - short and long measurement routes}
\label{fig_measurement_routes}
\end{figure}

The measurements have been carried out in various conditions to reflect better the true impact of the behavior of the communication channel and to average the impact of numerous short-term phenomena. What needs to be mentioned, the course of the route was selected to be representative. Thus it encompasses different route types, e.g., urban (including intersections) or highway (surrounded with, e.g., agricultural lands, sound-absorbing screens, forests, e.t.c). The whole measurement campaign was designed to resemble mostly the typical conditions of the V2V communications, e.g., small omnidirectional antennas placed directly on the cars' roofs, measurement route covering both motorways and urban area, varying distance and directions of utilized cars. This was to create a single and mostly universal channel model for this application. In particular, five runs (routes) have been identified, which are summarized in Tab.~\ref{tab:routes}, and graphically illustrated in Fig.~\ref{fig_measurement_routes}. 
Mainly, the first route (denoted in the figure by a black line and described as \textit{long route}) started near the premises of the Poznan University of Technology (PUT), close to the city center. It went over \SI{12}{\km} in the urban environment (with typical daily traffic conditions, with several traffic lights and some small traffic jams), reaching finally the high-speed road, where both cars drove over \SI{18}{\km}. The road consists of two two-lane streets separated by the green belt and passes through various regions. First, it went through the Poznan suburbs, then it crossed fields and woods, finally reaching the suburbs of the small town (Kórnik). The second route, denoted in Fig.~\ref{fig_measurement_routes} by the red dashed line and described as \textit{shorter route}, goes in the high-speed part of the first route, i.e., it overlaps with the first road over the last \SI{18}{\km}. Finally, there was a short route, denoted in the figure by a blue dotted line and described as \textit{opposite driving route}, where we have tested the impact of platoons driving in opposite directions. The length of this road was approximately equal to \SI{2}{\km}, and the crossing point was around the middle of this route. As the high-speed part of the routes went both over the woody and rural areas, it passed in some places close to the small clusters of houses. In such cases, these houses have been separated from the high-speed road by the so-called noise barriers (also called sound walls). 

The \textit{long route} has been followed once, whereas the shorter and the opposite-driving routes have been selected twice in order to collect a reasonable number of samples to make our analysis statistically tractable. Moreover, five runs were conducted in heavy/moderate traffic conditions during the daytime (around noon), when the number of cars on the street was high. In particular, heavy traffic was observed mainly in the urban area (i.e., there were some traffic jams along the route), and temporarily on the high-speed road (when so many cars were traveling together for some time that, e.g., the change of the lane was not possible for a longer time). By moderate traffic, we treat the situation where there are even several cars in the vicinity, but the car flow is not affected in any way, and most of the cars' maneuvers are possible in a reasonably short time. One run was carried out during the nighttime (around 10 p.m.) when the number of other cars on the road was minimal. The heavy-traffic measurements have been used for model derivation, whereas the low-traffic run for model validation, as reported in Sec.~\ref{sec:verification}. While diving into the urban environment, typical traffic situations have been observed, i.e., there were numerous traffic lights, some minor traffic jams, and cars changing the lanes. Moreover, it happened thrice that the transmitting car drove away while the receiver was blocked by the red light. Thus, all considered path loss scenarios (LOS, OLOS, NLOS) have been reflected while driving in the city. Similarly, all path loss conditions have been observed and carefully monitored in the highway measurements.

Next, to reliably detect the relationship between the observed path loss and distance between cars, the relative position of the transmitting and receiving cars changed during the whole measurement run. In the urban scenario, the distance between the cars was typically small, ranging between a couple of meters to some hundreds of meters. When focusing on the high-speed scenario, while guaranteeing the safety condition on the road, the minimum distance between the cars of around \SI{5}{\m} (bumper to bumper) has been achieved. Then, the distance was increased, reaching maximally around \SI{2}{\km}, and then decreased again to a few meters. Such a cycle has been repeated many times. Moreover, to get rid of the prospective impact of the velocity of the cars, the measurements have been conducted for various speeds. Mainly, as in the city center, the velocity varied between \SI[per-mode = fraction]{0}{\km\per\hour} to around \SI[per-mode = fraction]{50}{\km\per\hour}, while in the high-speed road, we have conducted the measurements in the following ways: 
\begin{itemize}
    \item first, in order to reflect typical car driving, the velocity was flexible and adjusted to the situation on the road; this approach has been applied for all roads beside 2 and 3 (see Tab.~\ref{tab:routes})
    \item second, as the platoons of trucks very often travel at a constant speed, we have fixed the car velocity to \SI[per-mode = fraction]{60}{\km\per\hour}, \SI[per-mode = fraction]{80}{\km\per\hour}, and \SI[per-mode = fraction]{100}{\km\per\hour} (of course, remembering that in various countries, there are various legal regulations regarding the speed of trucks); for each car speed, we have increased the distance from low (a few meters) to high (hundreds of meters), and decreased it back to the low one. 
\end{itemize}

\begin{table*}[h!tb]
\centering
\label{tab:routes}
\caption{Routes selected for conducting the measurement campaigns}
\begin{tabular}{|M{0.5cm}|m{5cm}|M{1.8cm}|M{1.5cm}|M{1.5cm}|M{1.8cm}|M{1.5cm}|}
\hline
\textbf{Run} & \centering \textbf{Description} & \textbf{Ref. to Fig.~\ref{fig_measurement_routes}} & \textbf{Total length [km]} & \textbf{Time} & \textbf{Traffic conditions} & \textbf{Direction of driving} \\ \hline
1 &  A route started at the PUT premises and went through the urban and suburban areas over the first \SI{12}{\km}. Next, the \SI{18}{\km} long high-speed road started, which went through suburbs, rural and woody areas. Finally, it reached the small town suburbs. & Long route (black solid line)& $2\times18$ & Around noon & Heavy traffic  (in urban area), moderate traffic & Same \\ \hline
2,3 & The \SI{18}{\km} long high-speed road started, that went through suburbs, rural and woody areas. Finally, it reached the small town suburbs. & Shorter route (red dashed line)& $2\times12$ & Around noon &  Heavy traffic / moderate traffic & Same \\ \hline
4 & As above & Shorter route (red dashed line)& $2\times12$ & Nighttime (10 p.m.) & Low traffic & Same \\ \hline
5,6 & \SI{2}{\km} long section of the high-speed route near a small town (Kórnik), where the cars drove in opposite directions on the parallel two-lane streets  & Opposite driving route (blue dotted line)& $1\times2$ & Around noon & Moderate traffic &  Opposite\\ \hline
\end{tabular}
\end{table*}

\section{Selection and Parametrization of Path Loss Model}
\begin{figure}[!t]
\centering
\includegraphics[width=3.4in]{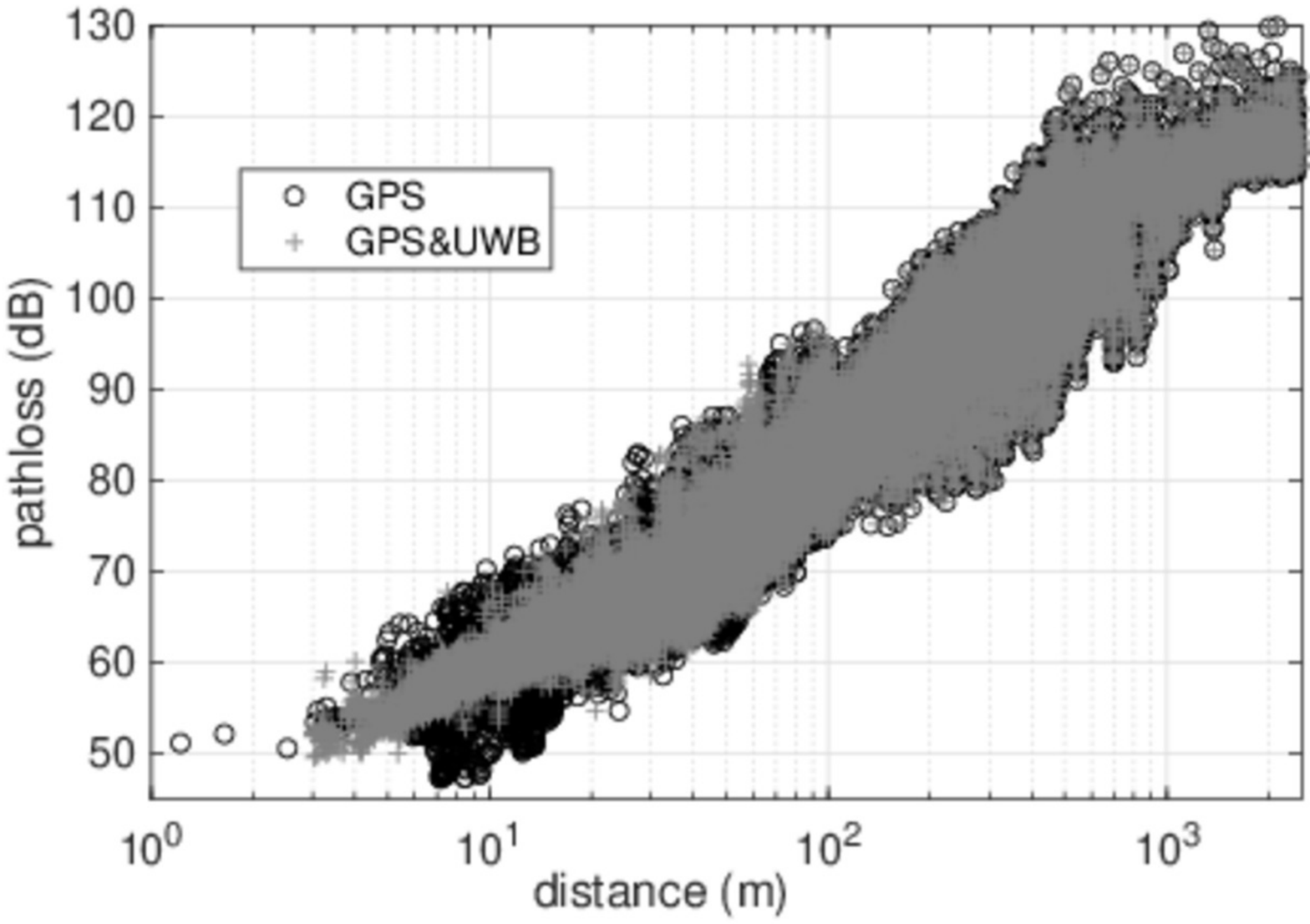}
\caption{Measured path loss as a function of TX-RX distance calculated using only GPS data and combination of GPS and UWB data.}
\label{fig_pathloss_vs_localization_method}
\end{figure}

The measured path loss as a function of distance is presented in Fig.\ref{fig_pathloss_vs_localization_method} for two distance estimation methods: GPS-only and GPS data supported by UWB measurements. More than 40000 measurement points are presented and used for modeling in the rest of this section. It is visible, as expected that the path loss increases with distance. Moreover, it is visible that the path loss above \SI{110}{\dB} can be erroneous due to an operation close to the sensitivity region of the test setup. This confirms that the censoring level of \SI{110.6}{\dB} calculated in Sec. \ref{sec_pathloss_est} meets the reasoning based on the plot. The last but crucial observation is about the accuracy of distance estimates for every measured point. While a few meter error possible in the case of GPS is not so important for distances higher than, e.g., \SI{100}{\m}, it significantly influences path loss modeling for shorter distances, the most relevant for intra-platoon communications. It is visible that UWB-supported measurements reduce the spread of path loss for a distance shorter than \SI{100}{\m}. GPS-only measurement also results in a single measurement of the inter-car distance of \SI{1.22}{\m}. This is physically impossible from the perspective of the antennas' locations on the car. For such kinds of measurements, the enhancement of GPS-based distance estimates is essential. From now on, only the UWB-enhanced distance data will be used. 

\subsection{Large-Scale Fading Modeling}
There are 4 hypothetical path loss and shadowing models considered in this work:
\subsubsection{Single slope model with lognormal shadowing}
\label{sec_model_SS}
Assuming the path loss at distance $d$ between TX and RX is denoted as $L(d)$ it should be equal to
\begin{equation}
    L(d)=L_{\mathrm{SS}}+10\gamma_{\mathrm{SS}}\mathrm{lg}\left(\frac{d}{10} \right)+X(0,\sigma_{\mathrm{SS}}^2),
    \label{eq:single_slope}
\end{equation}
where $\mathrm{lg}(~)$ denotes logarithm of base 10, $L_{\mathrm{SS}}$ is the reference path loss for 10 m distance, $\gamma_{\mathrm{SS}}$ is the path loss exponent and $X(0,\sigma_{\mathrm{SS}}^2)$ is a normal random variable of 0 mean and standard deviation $\sigma_{\mathrm{SS}}$. The last component represents shadowing, which typically is represented using a log-normal model when considering linear power. Hence, in this paper we use a Gaussian-distributed zero-mean random variable, following \cite{abbas2015measurement, Cheng_5_9GHz_Channel_model}, as logarithmic power is considered in the equations. This is the simplest model, but has been used previously to model V2V communications in \cite{Cheng_5_9GHz_Channel_model}. In total 3 coefficients have to be estimated.

\subsubsection{Single slope models with lognormal shadowing separate for LOS, OLOS and NLOS}
\label{sec_model_NLOS_OLOS}
In this case, the path loss is modeled independently for each type of the link, i.e.,
\begin{equation}
    L(d)\!\!=\!\!\begin{cases}
   \!\! L_{\mathrm{LO}}\!+\!10\gamma_{\mathrm{LO}}\mathrm{lg}\!\left( \frac{d}{10}\right)\!+\!\!X(0,\sigma_{\mathrm{LO}}^2) &\!\! \!\!\!\text{for LOS}\\
    \!\!L_{\mathrm{OLO}}\!+\!10\gamma_{\mathrm{OLO}}\mathrm{lg}\!\left( \frac{d}{10}\right)\!+\!\!X(0,\sigma_{\mathrm{OLO}}^2) &\!\! \!\!\! \text{for OLOS}\\
    \!\!L_{\mathrm{NLO}}\!+\!10\gamma_{\mathrm{NLO}}\mathrm{lg}\!\left( \frac{d}{10}\right)\!+\!\!X(0,\sigma_{\mathrm{NLO}}^2) &\!\! \!\!\! \text{for NLOS}
    \end{cases},
\end{equation}
where for each link type, separate coefficients are defined with meaning adequate to (\ref{eq:single_slope}). In total, nine coefficients have to be estimated. One should also note that although we distinguish here different propagation scenarios based on the existence of the LOS component, in every case, multipath propagation is observed, {\PK potentially prone to shadowing (blockage) of only selected propagation paths by, e.g., obstacles present or not within first Fresnel zone between TX and RX}. Thus, in each scenario, we account for such additional changes in attenuation using the $X(0,\sigma_{(.)}^2)$ parameter {\PK following, e.g., \cite{abbas2015measurement}.}

\subsubsection{Double slope model with single lognormal shadowing}
A model defined with two path loss exponents, i.e., $\gamma_{\mathrm{DSSS},1}$ and $\gamma_{\mathrm{DSSS},2}$ for distance below and above breakpoint distance $d^{\mathrm{DSSS}}_{\mathrm{thr}}$  denoted as Double Slope Single Shadowing (DSSS) is given by
\begin{equation}
\label{eq:modelDSS}
    L(d)\!\!=\!\!\begin{cases}
    \begin{aligned}
   \!\! &L_{\mathrm{DSSS}}\!+\!10\gamma_{\mathrm{DSSS},1}\mathrm{lg}\!\left( \frac{d}{10}\right) \\
   &+\!\!X(0,\sigma_{\mathrm{DSSS}}^2)
   \end{aligned}
   &\!\! \!\!\!\!\!\text{if } \!d<\!d^{\mathrm{DSSS}}_{\mathrm{thr}}
   \vspace{3mm}
   \\
 \!\!\!\begin{aligned} \!\! &L_{\mathrm{DSSS}}\!+\!10\gamma_{\mathrm{DSSS},1}\mathrm{lg}\!\left(
 \frac{d^{\mathrm{DSSS}}_{\mathrm{thr}}}{10}
 \right)\\
 &\!+\!10\gamma_{\mathrm{DSSS},2}\mathrm{lg}\!\left(\!\frac{d}{ d^{\mathrm{DSSS}}_{\mathrm{thr}}}\!\right)\!\!+\!\!X(0,\sigma_{\mathrm{DSSS}}^2)
 \end{aligned}
 &\!\!\!\!\! \!\!\!\text{if } d\!\geq\! d^{\mathrm{DSSS}}_{\mathrm{thr}}
    \end{cases}
\end{equation}
where a single log-normal shadowing variable is used of standard deviation $\sigma_{\mathrm{DSSS}}$. Here $L_{\mathrm{DSSS}}$ is the reference path loss for ten-meter distance. Similar model was proposed for \SI{5.9}{\GHz} band in \cite{Cheng_5_9GHz_Channel_model}. It requires 4 coefficients to be estimated along with setting the breakpoint distance $d^{\mathrm{DSSS}}_{\mathrm{thr}}$.

\subsubsection{Double slope model with double lognormal shadowing}
It can be expected that the shadowing because of cars, buildings, etc., can be more varying for longer links than for shorter. As such the DSSS model can be extended to consider separate shadowing standard deviations $\sigma_{\mathrm{DSDS},1}$ and $\sigma_{\mathrm{DSDS},2}$ below and above the breakpoint distance denoted as $d^{\mathrm{DSDS}}_{\mathrm{thr}}$, respectively, giving  \begin{equation}
\label{eq:modelDSDS}
   \!\! L(d)\!\!=\!\!\begin{cases}
   \begin{aligned}
   \!\!& L_{\mathrm{DSDS}}\!+\!10\gamma_{\mathrm{DSDS},1}\!\mathrm{lg}\!\left( \frac{d}{10}\right)\!
   \\
   &+\!\!X(0,\sigma_{\mathrm{DSDS},1}^2)
   \end{aligned}
   &\!\! \!\!\!\text{if } d\!<\!d^{\mathrm{DSDS}}_{\mathrm{thr}}
   
   \vspace{3mm}
   \\
 \!\!\!\begin{aligned} \!\! &L_{\mathrm{DSDS}}\!+\!10\gamma_{\mathrm{DSDS},1}\mathrm{lg}\!\left(\! \frac{d^{\mathrm{DSDS}}_{\mathrm{thr}}}{10}\!\right)\\
 &\!+\!10\gamma_{\mathrm{DSDS},2}\mathrm{lg}\!\left(\!\frac{d}{ d^{\mathrm{DSDS}}_{\mathrm{thr}}}\!\right)\!\!+\!\!X(0,\sigma_{\mathrm{DSDS},2}^2)
 \end{aligned}\!\!\!
 &\!\!\!\!\! \!\!\!\text{if } d\!\geq\! d^{\mathrm{DSDS}}_{\mathrm{thr}}
    \end{cases}.
\end{equation}
 DSDS stands for Double Slope, Double Shadowing. As a result of separate estimation using different degrees of freedom, it is expected that the reference path loss for \SI{10}{\m} distance $L_{\mathrm{DSDS}}$ and path loss exponents $\gamma_{\mathrm{DSDS},1}$ and $\gamma_{\mathrm{DSDS},1}$ will be different than in DSSS model.
The estimation of model parameters is carried out using the Maximum Likelihood (ML) estimator presented in \cite{Gustafson_ML_estimation_censoring_2015}. This takes into account that the path loss measurements above \SI{110.6}{\dB} are not reliable and are censored by the estimator. It prevents "flattening" of path loss vs. distance plot for high distances (visible in Fig. \ref{fig_pathloss_vs_localization_method}) caused by the limited dynamic range of the measurement setup to influence estimation results. There are approximately 9\% of censored samples in the dataset. The likelihood function is optimized using MATLAB, which finds the minimum negative logarithm of the likelihood function by adjusting large scale fading model parameters, e.g., path loss exponent or shadowing variance. As the estimated parameters in the model defined in Sec. \ref{sec_model_NLOS_OLOS} are specific only for a given link type, e.g., OLOS, there are three separate estimations conducted in this case. The resultant log-likelihood function values can be added, resulting in the total log-likelihood of this model. All four models are compared using Bayesian Information Criteria (BIC) \cite{Konishi2008}. It takes into account log-likelihood function values that are "penalized" using a number of measurement samples used for modeling and the number of parameters used by the tested large scale fading model, e.g., 3 in the case of the single slope model of Sec. \ref{sec_model_SS} and 9 in the case of the single slope model, separate for each link type as described in Sec. \ref{sec_model_NLOS_OLOS}. 

While the considered estimator overcomes the data censoring problem, another potential issue is an unequal number of samples available for each distance over the measurements range. It can be solved, e.g., by using the samples weighting method \cite{Karttunen_pathloss_2016}. Fortunately, because of properly carried measurements, the samples are quite equally distributed over distance, and usage of such a method is not needed here.

\begin{figure}[htb]
\centering
\includegraphics[width=3.0in]{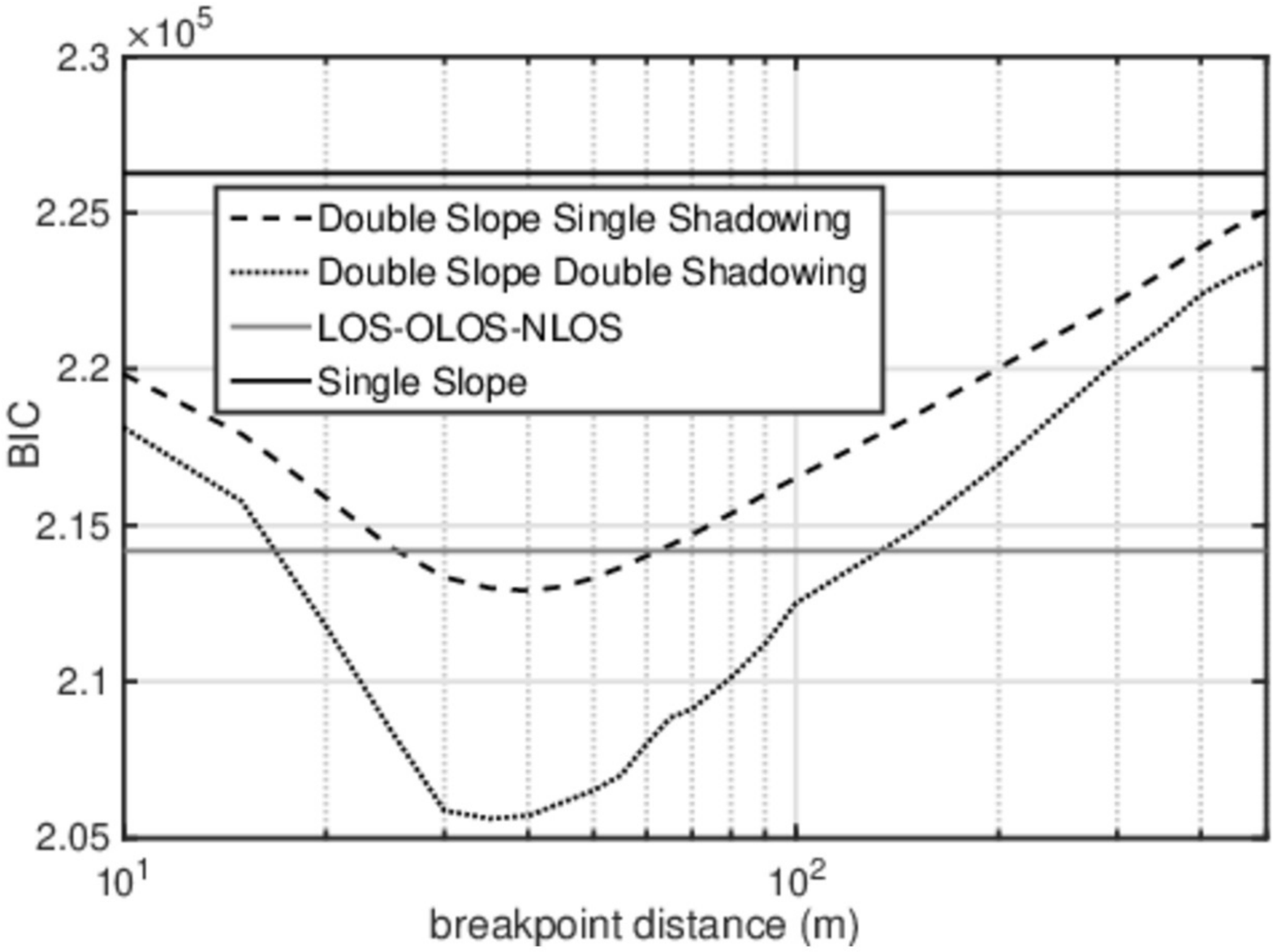}
\caption{Values of Bayesian Information Criteria (BIC) for all compared path loss models in function of breakpoint distance used by two models.}
\label{fig_BIC}
\end{figure}
As both considered double slope models need breakpoint distances to be specified, a set of values has been tested and compared using BIC value as shown in Fig. \ref{fig_BIC}. As a minimal BIC value suggests the best model fitting, it can be concluded that the model of choice should be the Double Slope Double Shadowing model with a breakpoint distance of 35 m.
As discussed in \cite{Cheng_5_9GHz_Channel_model,abbas2015measurement}, this value should be in theory the same as the distance of a link that results in the first Fresnel zone to touch the ground.{\PK The justification of path loss exponent change after distance exceeds the breakpoint distance is given in \cite{He_fresnel_zone_2012} for a two-path propagation scenario.}
A commonly used formula $\frac{4 h_{TX} h_{RX}}{\lambda}$ results in a theoretical breakpoint distance of \SI{29.6}{\m}. This is very close to the estimated value suggesting that both the measurements, especially distance measurements for short links, and estimation have been done correctly. 
Commonly used in the \SI{5.9}{\GHz} band LOS-OLOS-NLOS model seems to be not appropriate for the considered frequency band, probably because of the different nature of propagation at frequencies around ten times lower than used by DSRC transceivers. While the LOS-OLOS-NLOS model can inherently model attenuation introduced by a blocking car, the DSDS model does it statistically by a normal-distributed shadowing. The DSDS model has an additional advantage as it does not need to categorize the current link type. As mentioned in the manuscript’s introduction, the utilization of terrestrial television frequencies for V2V communications would probably need the utilization of a Radio Environment Map. While this entity cannot easily categorize link types, the suggested channel propagation model is highly advantageous due to its simplicity.

\begin{figure}[htb]
\centering
\includegraphics[width=3.2in]{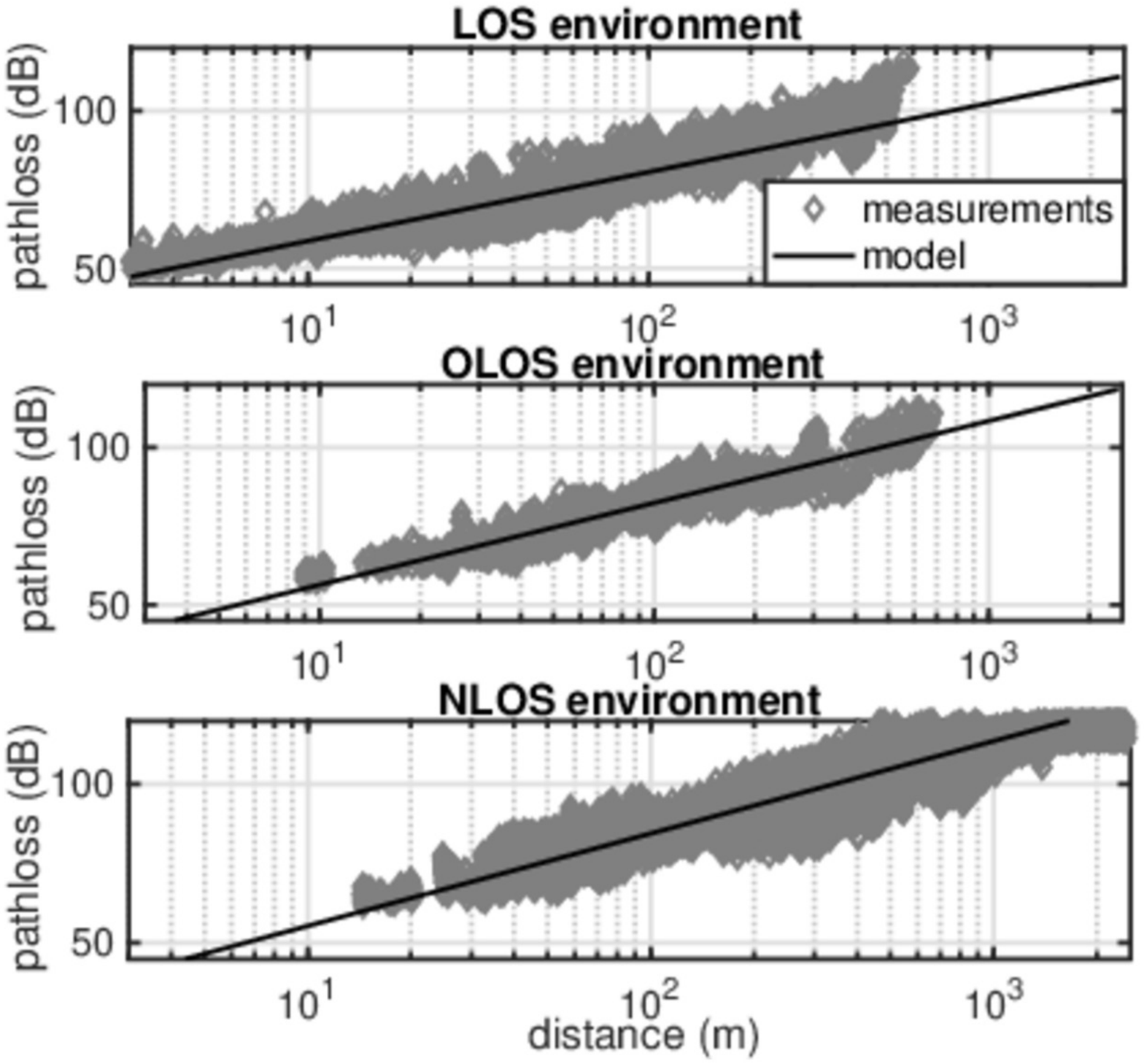}
\caption{Measured path loss and result of modeling with a single slope model separate for LOS, OLOS and NLOS environment.}
\label{fig_pathloss_LOS_NLOS_OLOS}
\end{figure}

\begin{figure}[tb]
\centering
\includegraphics[width=3.2in]{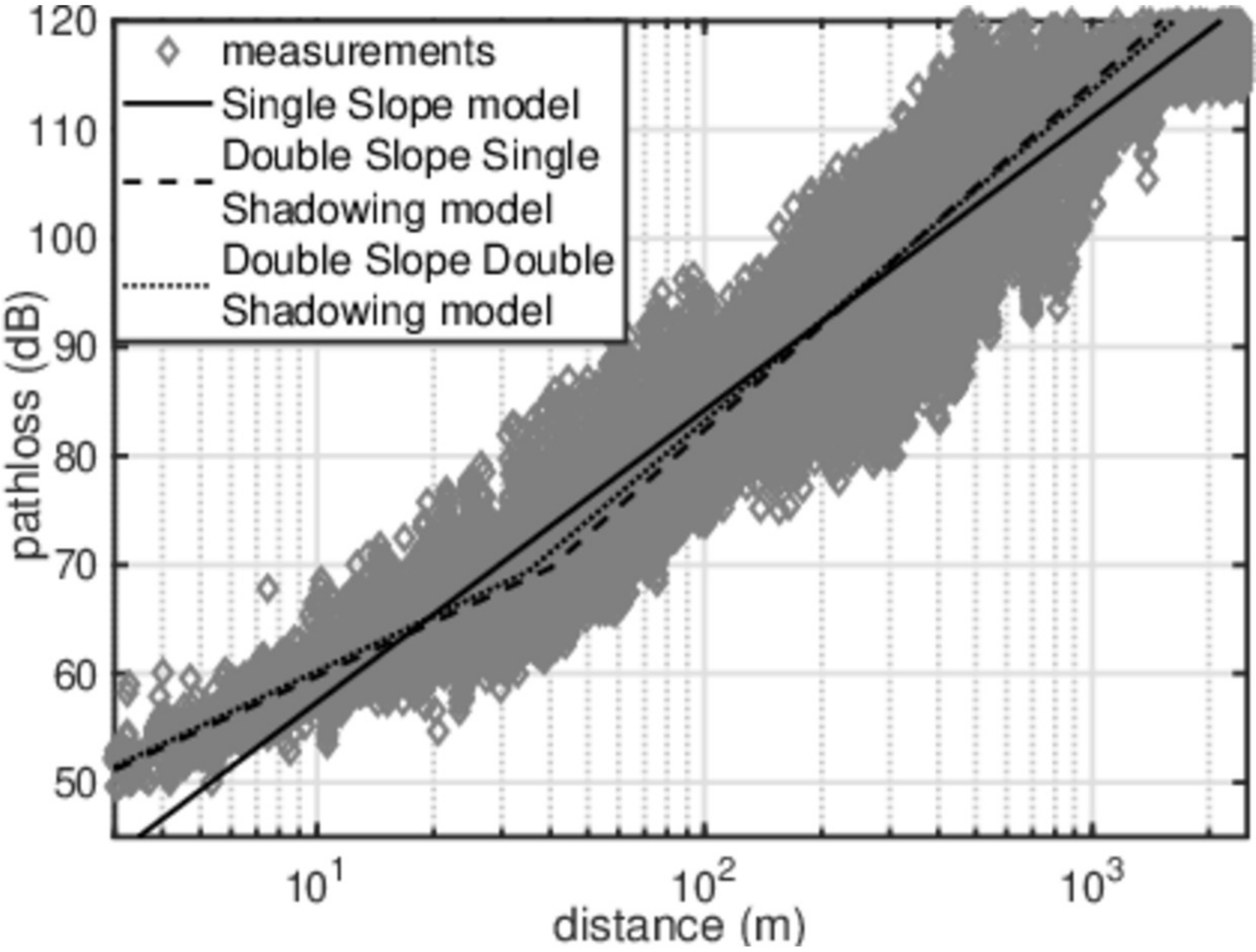}
\caption{Measured path loss and result of modeling with a single slope, and both considered double slope models.}
\label{fig_pathloss_vs_models}
\end{figure}

While the estimated parameters for each of considered models are presented in Table \ref{tab_model_par}, the modeled path losses as functions of distance are presented in Fig. \ref{fig_pathloss_LOS_NLOS_OLOS} and Fig. \ref{fig_pathloss_vs_models}. 
Moreover, a Root Mean Squared Error (RMSE) of each of the considered path loss models is calculated to be used as a fitting metric parallel to BIC. The RMSE equals \SI{4.45}{\dB}, \SI{3.95}{\dB}, \SI{3.72}{\dB}, and \SI{3.73}{\dB} for SS, LOS-OLOS-NLOS, DSSS, and DSDS models, respectively. These values reflect the order of BIC values with the exception for DSSS and DSDS models. The DSSS model results in a lower value of RMSE than the DSDS model. However, it has to be considered that the RMSE measures only fitting of the path loss model while the BIC considers how well both path loss and shadowing are fitted.
\begin{table}[tbh]
\caption{ML estimated parameters of all considered path loss models}
\label{tab_model_par}
\centering
\begin{tabu}{|p{0.8\linewidth}|}
\hline
\multicolumn{1}{|c|}{\textbf{Single Slope}}                  \\ \hline
$L_{\mathrm{SS}}=57.34.44 dB$; $\gamma_{\mathrm{SS}}=2.69$; $\sigma_{\mathrm{SS}}=4.5dB$ \\ \hline
\multicolumn{1}{|c|}{\textbf{LOS-OLOS-NLOS}}                 \\ \hline
$L_{\mathrm{LO}}=58.6 dB$; $\gamma_{\mathrm{LO}}=2.19$; $\sigma_{\mathrm{LO}}=3.3 dB$;\linebreak
$L_{\mathrm{OLO}}=56.2 dB$; $\gamma_{\mathrm{OLO}}=2.6$; $\sigma_{\mathrm{OLO}}=3.7 dB$; \linebreak
$L_{\mathrm{NLO}}=55.3 dB$; $\gamma_{\mathrm{NLO}}=2.91$; $\sigma_{\mathrm{NLO}}=4.9 dB$; 
\\ \hline
\multicolumn{1}{|c|}{\textbf{Double Slope Single Shadowing}} \\ \hline
$d^{\mathrm{DSSS}}_{\mathrm{thr}}=40m$;
$L_{\mathrm{DSSS}}=59.7 dB$; $\gamma_{\mathrm{DSSS},1}=1.65$; $\gamma_{\mathrm{DSSS},2}=3.19$;
$\sigma_{\mathrm{DSSS},1}=3.85 dB$;
\\ \hline
\multicolumn{1}{|c|}{\textbf{Double Slope Double Shadowing}} \\ \hline
$d^{\mathrm{DSDS}}_{\mathrm{thr}}=35 m$; $L_{\mathrm{DSDS}}=59.8 dB$; $\gamma_{\mathrm{DSDS},1}=1.6$; $\gamma_{\mathrm{DSDS},2}=3.14$;
$\sigma_{\mathrm{DSDS},1}=2.2 dB$;
$\sigma_{\mathrm{DSDS},2}=4.5 dB$;
\\ \hline
\end{tabu}
\end{table}

In the case of the LOS-OLOS-NLOS model, it is visible in Fig. \ref{fig_pathloss_LOS_NLOS_OLOS} that the measurements for the LOS link are typically obtained for shorter TX-RX distances. This is an expected phenomenon, similarly to an increase of path loss exponent, the more "blocked" is the link. Similarly, the link environment increases shadowing standard deviation from \SI{3.3}{\dB} for LOS up to \SI{4.9}{\dB} for NLOS. Unfortunately, the reference path loss for ten-meter distance takes counter-intuitive values. It is the highest for the LOS link, while it is expected that OLOS or NLOS propagation will result in additional attenuation due to blockage by cars or buildings. This expectation is confirmed by the measurements carried for the \SI{5.9}{\GHz} band \cite{abbas2015measurement}. This suggests again that the propagation at a frequency nearly ten times smaller than \SI{5.9}{\GHz} has other driving effects and should be modeled differently.     

\begin{figure}[htb]
\centering
\includegraphics[width=3.2in]{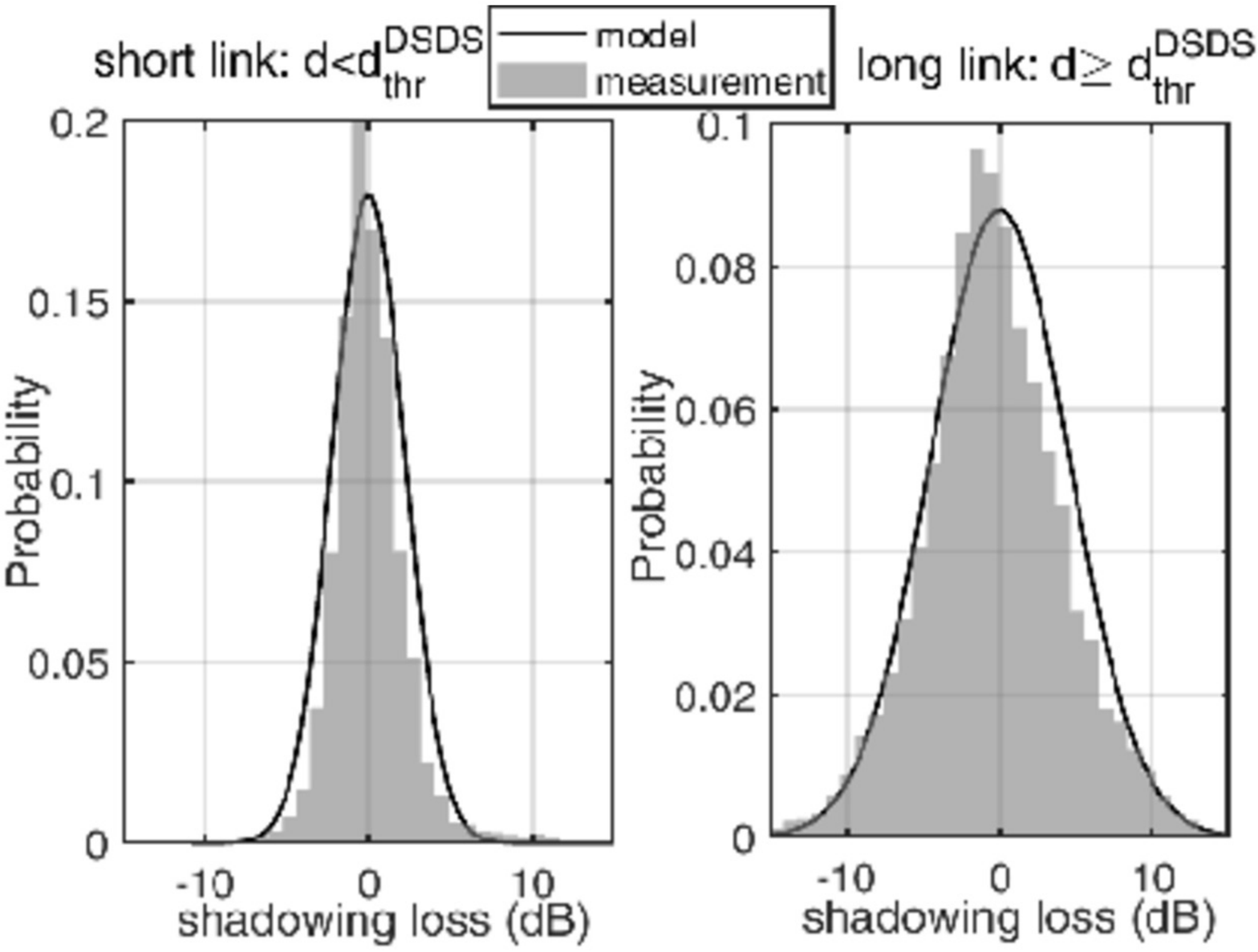}
\caption{Probability distribution of shadowing samples for DSDS model: measured and modeled using Gaussian probability density function; Separately for short and long links.}
\label{fig_shadowing_dist}
\end{figure}

The advantages of double slope models over a single slope model are visible in Fig. \ref{fig_pathloss_vs_models}. It is visible that the single slope model underestimates path loss for short distances and also seems to underestimate it for the longest links. At the same time, double slope models are able to reflect the path loss change with distance better. Both DSDS and DSSS models have very similar coefficients except for shadowing standard deviations. In the case of the DSDS model, the path loss exponent for short links equals 1.6. This value is lower than the theoretical value of 2 for Line-of-Sight propagation. While this is probably the effect of two-ray propagation, the fadings are not visible in the measured data. For the \SI{5.9}{\GHz} band, the two-ray model is commonly used~\cite{Boban_TVT_2014} for short links as a result of its visible appearance in measurement results \cite{Nilsson_2017_V2V_model}. For links longer than the breakpoint distance, the path loss exponent rapidly increases for the DSDS model.
Similarly, the shadowing standard deviation is much higher for longer links. 
In order to further analyze the shadowing of the DSDS model, histograms of measured values are presented in Fig.~\ref{fig_shadowing_dist} along with Gaussian Probability Density Functions (PDFs) resulting from modeling. It is visible that the histogram is reflected quite well by the Gaussian distribution both for short ($d<d^{\mathrm{DSDS}}_{thr}$) and long ($d\geq d^{\mathrm{DSDS}}_{thr}$) links. Some errors (i.e., the difference between the models and measured shadowing) can be the effect of censored samples, in the case of a long link, and two ray propagation, in the case of a short link. However, the Gaussian random variable should allow for simple yet precise enough modeling of the shadowing effect in the considered use case.

\begin{figure}[tb]
\centering
\includegraphics[width=3.1in]{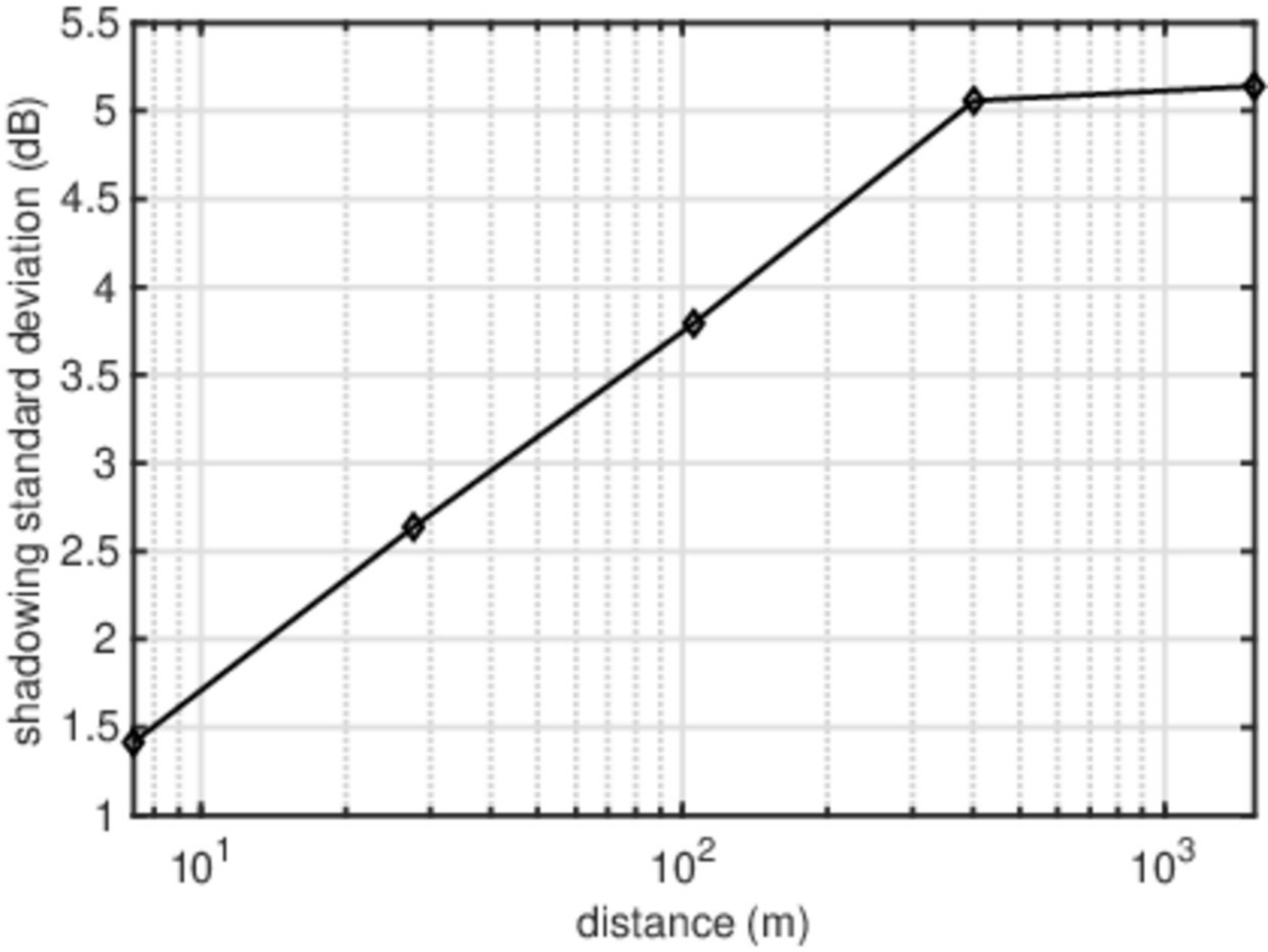}
\caption{ Variation of the shadowing standard deviation as function of distance between cars.}
\label{shadowing_vs_distance_for_DSDS}
\end{figure}

We have also verified how shadowing varies as a function of the distance between the transmitter and the receiver in the DSDS case. From all measurement samples, the achieved path loss values have been subtracted, and the resultant values have been split into five sets corresponding to five TX-RX distance ranges. For each set, the shadowing standard deviation has been calculated. As shown in Fig.~\ref{shadowing_vs_distance_for_DSDS}, the value of shadowing standard deviation in the logarithmic scale increases linearly to reach the plateau. The standard deviation becomes stable when the distance between the cars is above a certain threshold (around \SI{400}{\m}). 

\subsection{Shadowing Auto-Correlation}
An important aspect, especially for simulation of V2V communications in continuous time, is the correlation of shadowing samples in time or/and space. As suggested in \cite{Nilsson_2017_V2V_model}, typical measurements do not allow to formulate a reliable statistical model. The main problem comes from the simultaneous movement of the TX, the RX, and possibly other elements of the environment. Typically, a Gudmundson model \cite{Gudmundson_correlation_1991} is used, e.g., in \cite{abbas2015measurement,Nilsson_2017_V2V_model}, as it is relatively simple yet well established for wireless communications. 

Let us define $x_i$ as $i$-th measured shadowing sample obtained at time $t_i$ and at distance $d_i$. While the time can be just an absolute time value, e.g., Universal Absolute Time, the definition/type of distance is rather unspecified. It can be a distance of travel of the TX, a distance of travel of the RX, or their combination considering in addition, e.g., the change in distance between TX and RX. While \cite{abbas2015measurement} does not provide a definition of distance, \cite{Nilsson_2017_V2V_model} uses the cumulative distance of travel using the mean of instantaneous speed of TX and RX. On the other hand, in \cite{Gustafson_2020} authors utilize the sum of distance traveled by TX and RX. In all the cases, autocorrelation is calculated as
\begin{equation}
    R(\Delta d)=\frac{1}{N(\Delta d)}\sum_{i,j: |d_i - d_j|= \Delta d } x_{i} x_{j},
    \label{eq_autocorr_dis}
\end{equation}
where $N(\Delta d)$ is a number of elements summed for a given value of $\Delta d$ being a distance difference. Mathematically it is a cardinality/size of set $\left(i,j: |d_i - d_j|= \Delta d \right)$. The above equation correlates all shadowing samples $x_{i}$ that were obtained at points distanced by $\Delta d$. The factor in front of the sum divides the sum by the number of summed elements. Obviously, for $\Delta d=0$ the variance of shadowing should be obtained, i.e., $\sigma^{2}$. 
The Gudmundson model assumes that for the decorrelation distance $d_c$, the autocorrelation can be presented as
\begin{equation}
    R(\Delta d)=\sigma^{2} e^{-\frac{\Delta d}{d_c}}.
\end{equation}

\begin{figure}[htb]
\centering
\includegraphics[width=3.1in]{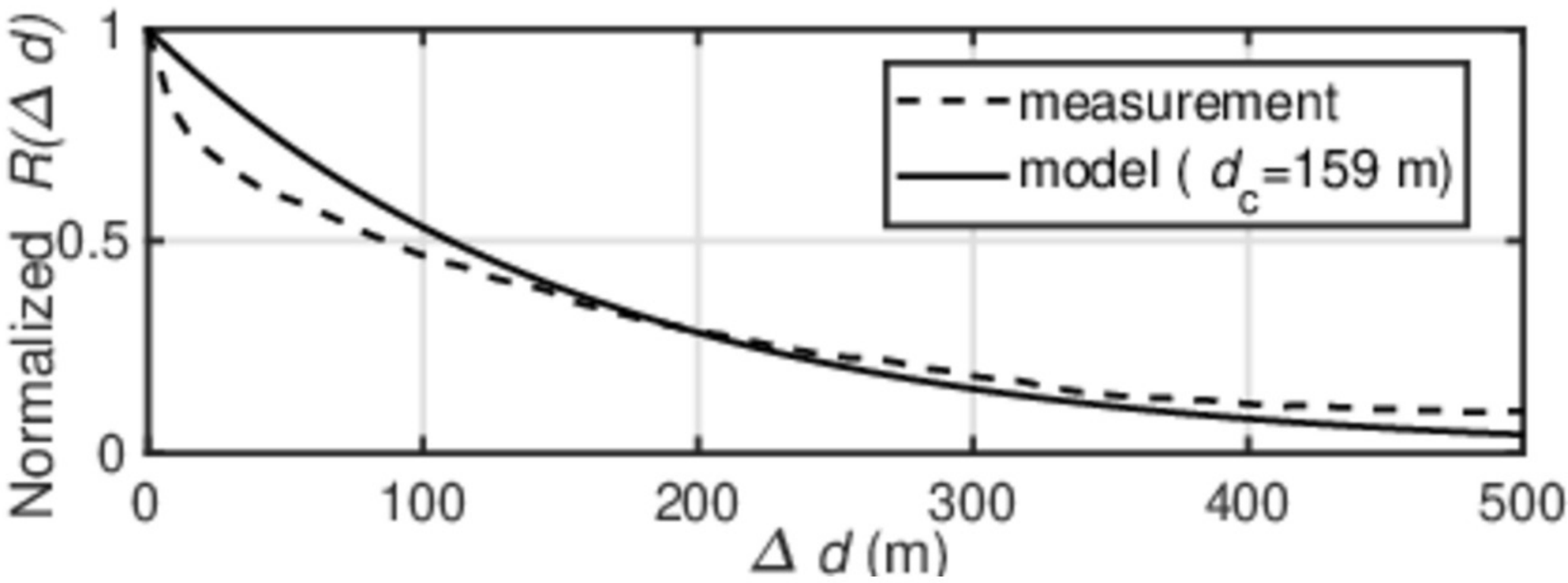}
\caption{Normalized shadowing correlation as a function of change of distance traveled by RX car. All samples considered.}
\label{fig_shadowing_corr_dist_all}
\end{figure}

In order to process shadowing samples obtained from measurements first the censored samples were removed. Next, the data was divided into blocks, separated by periods of censored samples. Values of $d_i$ were calculated as a cumulative travel distance for the RX considering instantaneous speed reported by GPS $v_{\mathrm{RX},i}$, i.e., $d_i=d_{i-1} +|v_{\mathrm{RX},i}|(t_i-t_{i-1})$. Next, as the inter-measurement period, i.e., $t_i-t_{i-1}$, varies slightly, and inter-measurement distance, i.e., $d_i-d_{i-1}$, can vary significantly as a result of vehicles speed change, resampling was carried to obtain the same number of equally-distanced subsequent samples. This enables efficient digital implementation of (\ref{eq_autocorr_dis}) with $\Delta d$ being a discrete variable. The correlation is calculated using (\ref{eq_autocorr_dis}). Its shape after normalization (division by $\sigma^{2}$) is shown in Fig. \ref{fig_shadowing_corr_dist_all}. The Gudmundson model is fitted using weighted least squares:
\begin{equation}
    \min_{d_c} \sum_{\Delta d}N(\Delta d)\left(\frac{R(\Delta d)}{\sigma^{2}} -e^{-\frac{\Delta d}{d_c}}\right)^{2}
\end{equation}
to take into account the reduced variance of $R(\Delta d)$ for small values of $\Delta d$ as a result of increased number of $x_{i}x_{j}$ products. It is visible that the model matches measurements relatively well with an estimated $d_c$ equal to \SI{159}{\m}. However, while this calculation takes into account the movement of one car it does not consider the movement of the other car and neighboring vehicles. As such further analysis has been carried.

\begin{figure}[htb]
\centering
\includegraphics[width=3.1in]{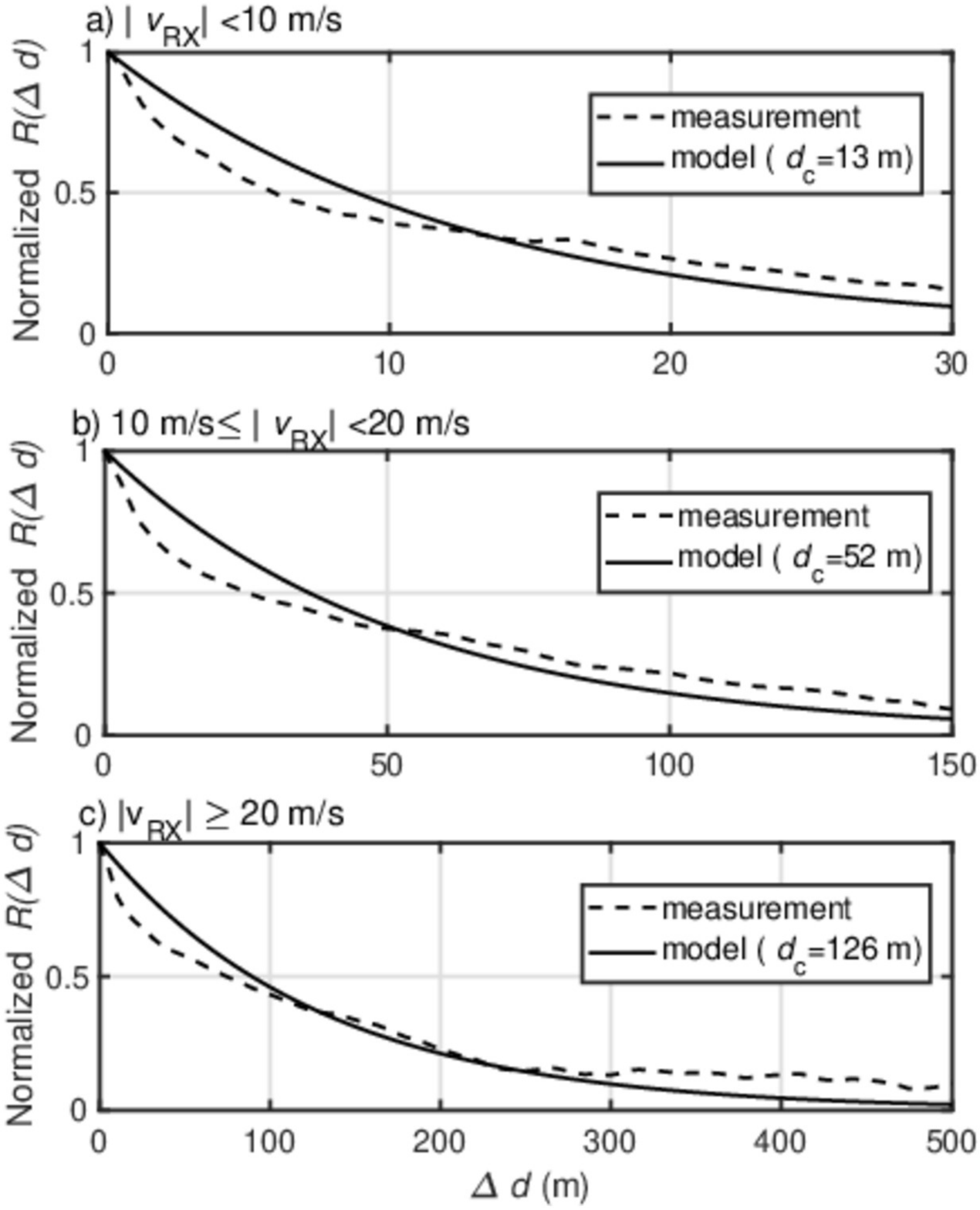}
\caption{Normalized shadowing correlation as a function of change of distance traveled by RX car. Only periods of $|v_{TX-RX}| < 5\frac{\mathrm{m}}{\mathrm{s}}$ considered. }
\label{fig_shadowing_corr_dist}
\end{figure}
\begin{figure}[tb]
\centering
\includegraphics[width=3.1in]{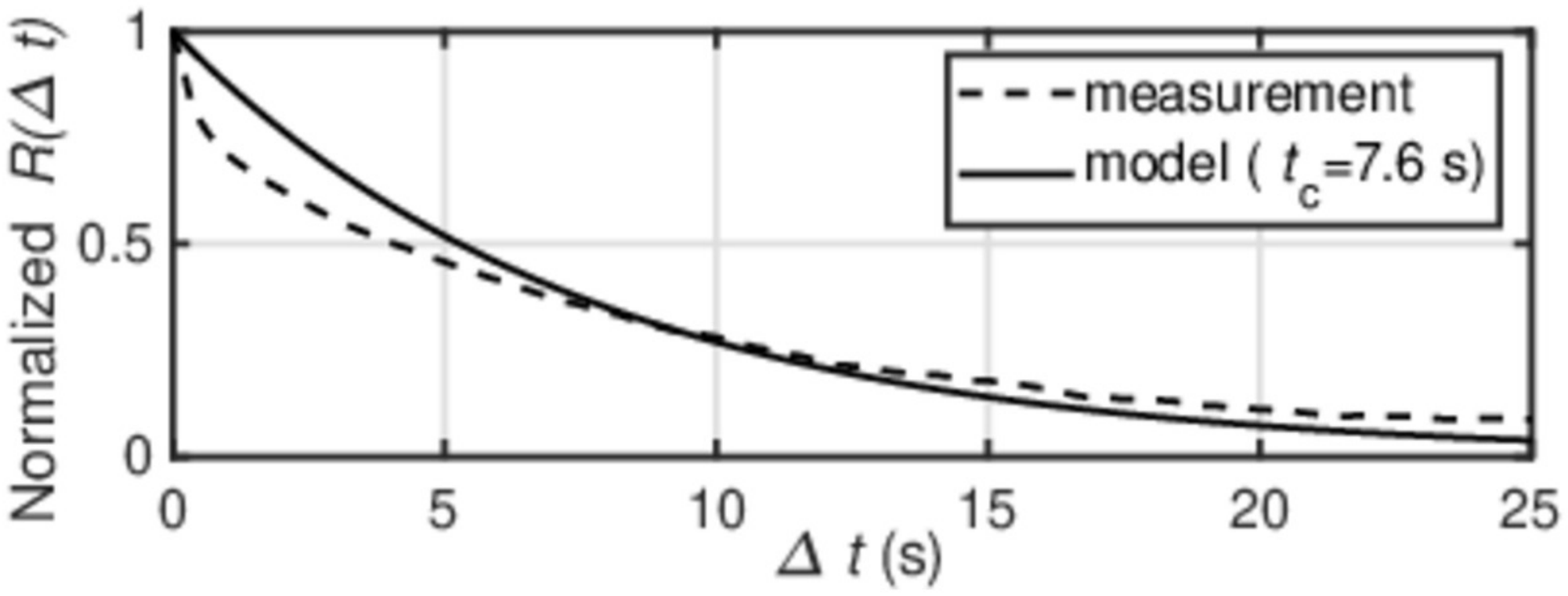}
\caption{Normalized shadowing correlation as a function of time change. All samples considered.}
\label{fig_shadowing_corr_time_all}
\end{figure}

The obtained shadowing measurements are classified according to the speed of one car, typically RX, and the relative speed of TX and RX cars denoted as $v_{\mathrm{TR-RX}}$. The shadowing autocorrelation functions for a subset of measurements when both cars are moving at a similar speed, i.e., $|v_{\mathrm{TR-RX}}| < 5\frac{\mathrm{m}}{\mathrm{s}}$, for three different ranges of RX car speeds, are shown in Fig.~\ref{fig_shadowing_corr_dist}. This should be, in most cases, the traffic pattern when both cars move in a platooning fashion. While in all the cases the Gudmundson model quite reliably reflects the shape of the autocorrelation function, the estimated decorrelation distance increases rapidly from \SIrange{13}{126}{\m} with increasing RX speed. The reason for such behavior is that the distance-based model accounts for velocity changes of a single vehicle (or rather changes in the relative velocity between the transmitter and the receiver), while both the transmitter and the receiver experience different variations with other surrounding objects also moving. As such, usage of a single $d_c$ value can result in significant errors in propagation modeling. A remedy to this problem may be to account for the movement of both - the transmitter and the receiver - separately, considering time as the parameter for the correlation calculation. This is justified by the increase of decorrelation distance with speed. It can be divided by speed, thus reducing the variations using time as a reference variable. Therefore, the autocorrelation is calculated as
\begin{equation}
    R(\Delta t)=\frac{1}{N(\Delta t)}\sum_{i,j: |t_i - t_j|= \Delta t } x_{i} x{j}
    \label{eq_autocorr_time}
\end{equation}
and the time-domain Gudmundson model is defined as 
\begin{equation}
    R(\Delta t)=\sigma^{2} e^{-\frac{\Delta t}{t_c}},
\end{equation}
where $t_c$ is the decorrelation time. 
\begin{figure}[!b]
\centering
\includegraphics[width=3.1in]{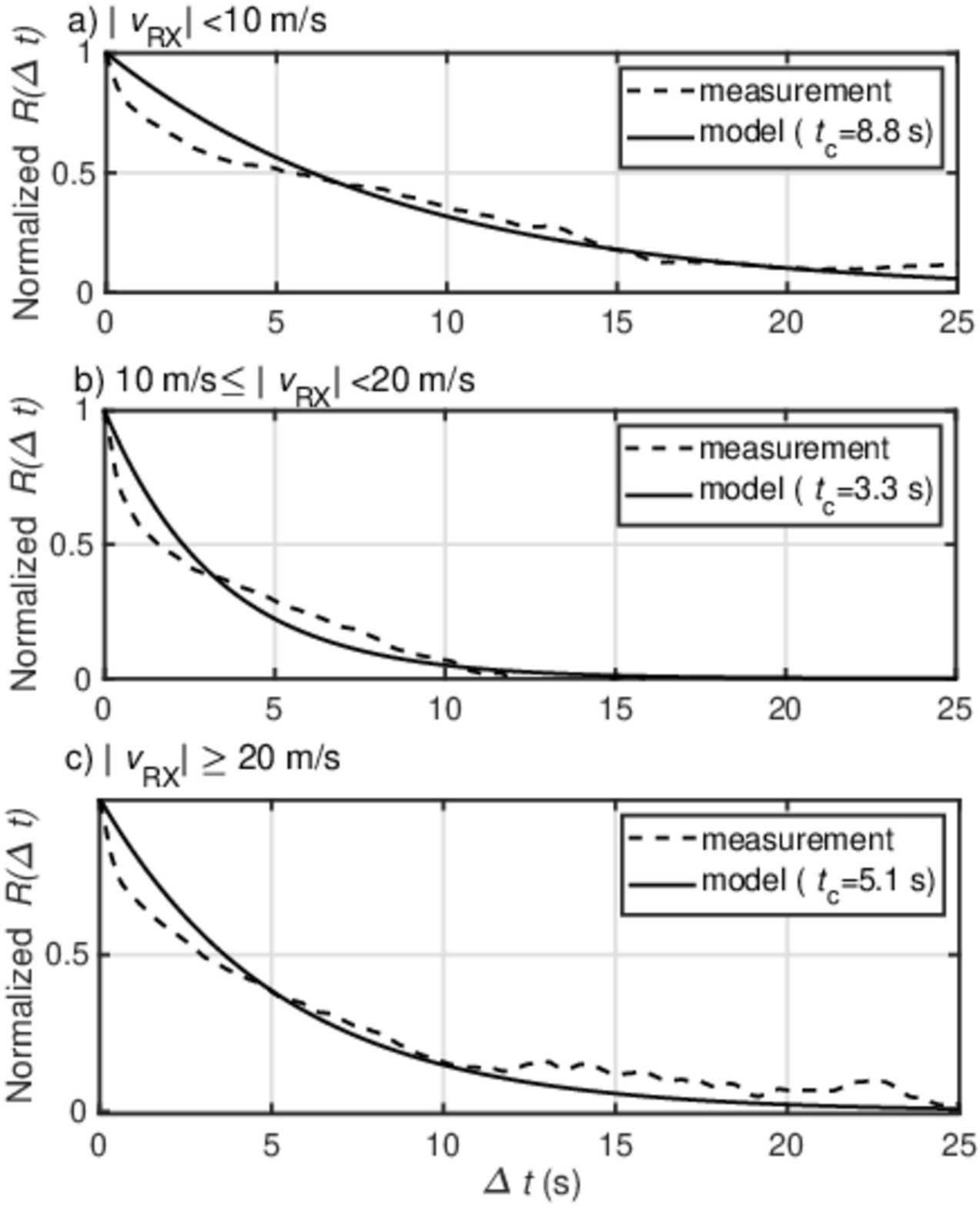}
\caption{Normalized shadowing correlation as a function of time change. Only periods of $|v_{\mathrm{TR-RX}}| < 5\frac{\mathrm{m}}{\mathrm{s}}$ considered. }
\label{fig_shadowing_corr_time}
\end{figure}
The result of autocorrelation calculations along with the fitted model for the same range of vehicle speed as in Fig. \ref{fig_shadowing_corr_dist} is shown in Fig. \ref{fig_shadowing_corr_time}. While the alignment of sample-based autocorrelation and model-based autocorrelation looks similar as previously, the decorrelation time ranges from \SIrange{3.3}{8.8}{\s}. While this might not be sufficiently accurate for some applications, the range of model parameter has been significantly reduced. The decorrelation time has been estimated for a number of speed ranges, as presented in Table \ref{tab_decorr_time}. As the estimated parameter values are split into different cases, the granularity of such division cannot be too fine as this will result in a small number of samples used to obtain the autocorrelation, which will result in low accuracy of the estimated $t_c$ value. Here, the results obtained with less than 100 samples at the level of $R(\Delta t)=\exp(-1)$ are not considered. This is marked as "$-$" in the table. The decorrelation time is in the range from \SIrange{3.3}{9.6}{\s}. It suggests a need for further measurements and modeling, as no relation describing the dependence of this parameter on the velocities can be found, and the values seem random. Another analysis, not presented here, also did not reveal the dependence of the decorrelation time on the sign of $v_{\mathrm{TX-RX}}$, i.e., cars approaching or moving away, or the distance between cars, i.e., above or below $d^{\mathrm{DSDS}}_{\mathrm{thr}}$. Therefore, considering the limited set of estimated $t_c$ values and no relation found between $t_c$ and velocities, a simple solution is to use a constant decorrelation time of \SI{7.6}{\s}. This is a result of $t_c$ obtained after analysis of all shadowing samples with the normalized autocorrelation function shown in Fig. \ref{fig_shadowing_corr_time_all}. 
\begin{table}[tb]
\caption{Estimated decorrelation time (in s) depending on $v_{\mathrm{RX}}$ and $v_{\mathrm{TX-RX}}$ range}
\label{tab_decorr_time}
\centering
\renewcommand{\arraystretch}{1.5}
\begin{tabular}{cl|r|r|}
\cline{1-4} 
\multicolumn{2}{|l|}{$|v_{\mathrm{TX-RX}}|$ (m/s)}                                   
& $\langle 0, 5)$    & $\langle 5, 15)$    \\ \hline
\multicolumn{1}{|c|}{\multirow{4}{*}{
\rotatebox[origin=c]{90}{$|v_{\mathrm{RX}}|$ (m/s)}
}} & $[ 0, 10)$  & 8.8                & -                   \\ \cline{2-4} 
\multicolumn{1}{|c|}{}                                  & $[ 10, 20)$ & 3.3                & 5.9                 \\ \cline{2-4} 
\multicolumn{1}{|c|}{}                                  & $[ 20, 30)$ & 5.8                & 9.6                 \\ \cline{2-4} 
\multicolumn{1}{|c|}{}                                  & $[ 30, 40)$ & 3.7                & 8.6                 \\ \hline
\end{tabular}
\end{table}

\section{Validation of the Large-Scale Fading Model by Another Set of Measurements}
\label{sec:verification}
Another measurement campaign was carried out three months after the initial one with a similar measurement setup and scenarios. The measurement has been carried out around 10 PM so that the influence of traffic intensity on the large-scale fading model should be visible if exists. The assumptions about the TX-RX movement pattern were the same as described in Sec.\ref{sec_meas_campaign} for the previous measurements. Though, instantaneous conditions, e.g., speed and the distance between cars, at a specific location were independent of the ones used in previous measurements. The only difference in measurement setup is the utilization of a band-pass filter Minicircuits ZABP-598-S+ at an input of spectrum analyzer to attenuate strong signals outside of the measurement band, reducing the overload probability. The input attenuation was set to 0 dB, reducing the internal noise floor and increasing the maximal reliably measured path loss threshold to \SI{123.5}{\dB}. This is used as the censoring level here. As only the parameters of the DSDS model were to be estimated, there was no need to classify instantaneous link as LOS/OLOS/NLOS. 

\begin{figure}[tb]
\centering
\includegraphics[width=3.2in]{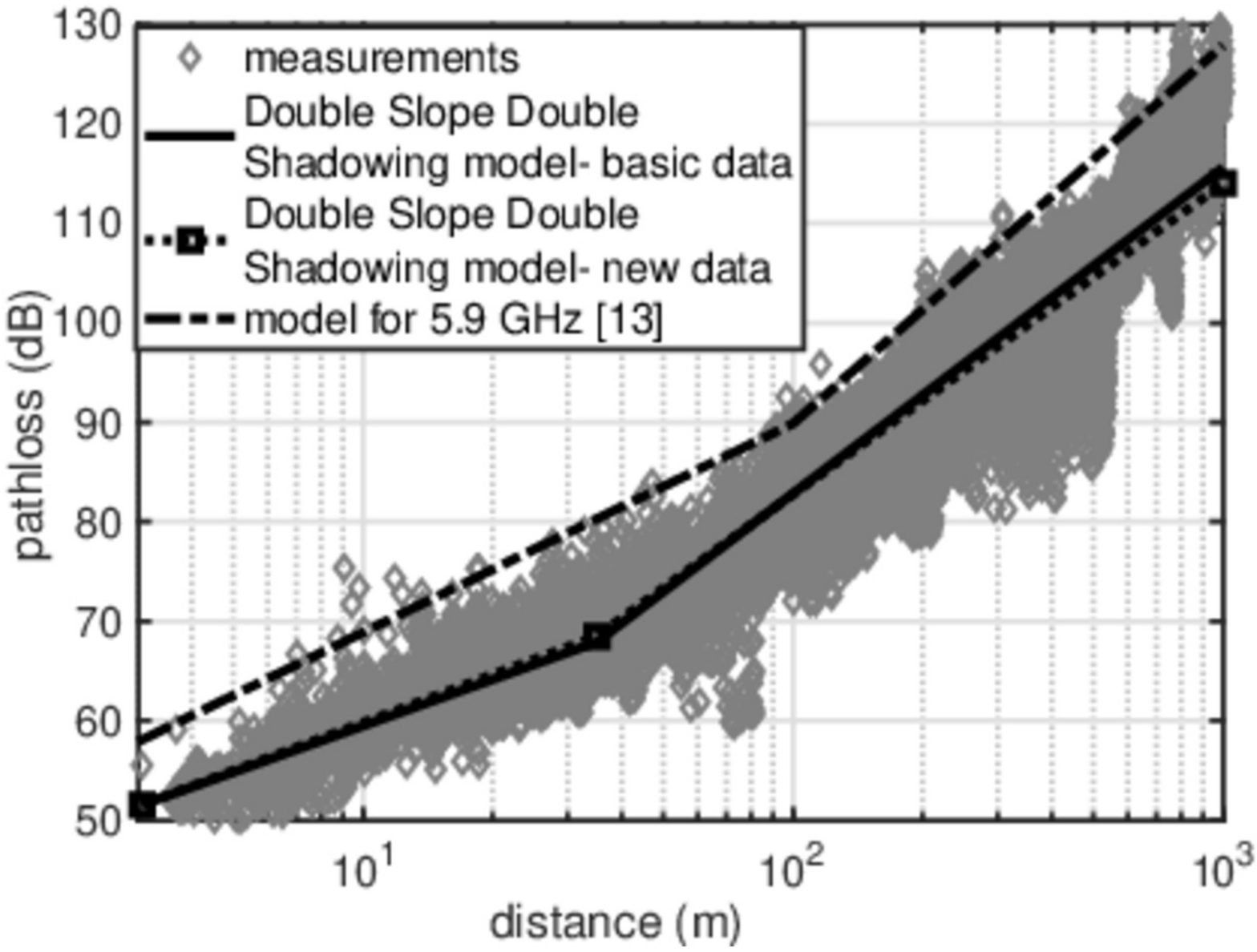}
\caption{Measured path loss and result of modeling with DSDS against DSDS model obtained for first measurement campaign and double slope model for \SI{5.9}{\GHz} band shown in \cite{Cheng_5_9GHz_Channel_model}}.
\label{fig_pathloss_vs_models_second_meas}
\end{figure}

The path loss measurements as a function of distance are shown in Fig. \ref{fig_pathloss_vs_models_second_meas}. The data were used for ML estimation of DSDS parameters assuming breakpoint distance equals \SI{35}{\m}. We keep this value the same as previously, as justified by the Fresnel zone model and prior measurements. The obtained parameters are $L_{\mathrm{DSDS}}=59.5$\si{\dB}, $\gamma_{\mathrm{DSDS},1}=1.55$, $\gamma_{\mathrm{DSDS},2}=3.28$, $\sigma_{\mathrm{DSDS},1}=2.2$\si{\dB}, $\sigma_{\mathrm{DSDS},2}=4.8$\si{\dB}. These are very close to the results for the first measurements campaign reported in Tab.~\ref{tab_model_par}. This observation can be confirmed by a graphical comparison of both models presented with dotted lines in Fig.~\ref{fig_pathloss_vs_models_second_meas}. Moreover, the decorrelation time equals $t_c=9.3$\si{\s}, which is reasonably close to the value obtained after the first measurements. This allows us to formulate a thesis that the presented DSDS model is suitable for modeling path loss (with shadowing) with significantly varying traffic conditions.

 Moreover, a double slope path loss model for propagation at \SI{5.9}{\GHz} band \cite{Cheng_5_9GHz_Channel_model} is added to Fig. \ref{fig_pathloss_vs_models_second_meas} for comparison. While the authors of \cite{Cheng_5_9GHz_Channel_model} did not provide the reference path loss, i.e., $L_{\mathrm{DSS}}$, in their work, here, it is calculated using the free-space propagation formula. While the reference path loss is probably underestimated, still the model for the \SI{5.9}{\GHz} band shows at least \SI{6}{\dB} higher path loss than the proposed DSDS model for TVWS. Moreover, the breakpoint distance equals \SI{100}{\m} at \SI{5.9}{\GHz}, being significantly higher than \SI{35}{\m}, optimal for TVWS. {\PK This results from the frequency-dependent first Fresnel zone size \cite{He_fresnel_zone_2012}.} The model obtained for the \SI{5.9}{\GHz} band is not valid in the TVWS. {\PK However, both models have similar statistics of shadowing. The model in \cite{Cheng_5_9GHz_Channel_model} uses standard deviation of \SI{2.6}{\dB} and \SI{4.4}{\dB} for the first and the second slope, respectively. These values are relatively close to the ones obtained by us for terrestrial TV bands, i.e., \SI{2.2}{\dB} and \SI{4.5}{\dB}, respectively.}

\section{Conclusions}
In this paper, the path loss model for vehicles platooning operating in TV bands is proposed, based on the extensive real-world measurements and considering the effect of shadowing auto-correlation. As four separate approaches have been investigated (mainly single slope with unique lognormal shadowing; single slope with separate lognormal shadowing for LOS, OLOS, and NLOS; double slope with single lognormal shadowing, and finally, double slope with double lognormal shadowing), finally the model of choice should be the Double Slope Double Shadowing model with breakpoint distance of \SI{35}{\m}. It has been proved that it achieves the best adjustment based on the Bayesian Information Criteria.
Moreover, it has been validated to work properly to some extent in various traffic conditions, i.e., 
the model derived for normal-day traffic mode can be easily applied to the night traffic mode. Moreover, despite the path loss modeling, the space-time autocorrelation of the shadowing is also considered. It has been derived that - considering the whole set of samples - the correlation distance equals approximately \SI{160}{\m}, whereas the correlation time - around \SI{7.6}{\s}. Clearly, these results are strictly related to cars' absolute and relative speed. 
The proposed path loss model (with shadowing) is significantly different from those proposed for V2V communications in the \SI{5.9}{\GHz} band. This results from a change of properties of primary propagation effects, e.g., diffraction, with approximately ten times wavelength shortening. Most importantly, the TV band seems to be advantageous for V2V communications, considering the signal blockage by cars is statistically insignificant, and the valid model parameters do not change significantly in time.  The model can be used to estimate path loss as a function of V2V distance with lognormal distributed shadowing. The simplicity of the final model allows for its simple implementation for V2V systems design and simulations. 
 Finally, let us highlight the selected prospective investigation directions that will further improve the proposed model. First, detailed studies on the applicability of the proposed model to various frequencies have to be done; it is necessary to answer if any correction factors are necessary when dealing with distant central frequencies. Moreover, it will be beneficial to investigate the impact of the montage place of the transmit/receive antenna on the propagation properties; in our experiment, the antennas have been placed on the rooftop, probably other phenomena will appear when the antennas are mounted at the bumper level or even below cars. Finally, it will be interesting to evaluate the precise impact of the surrounding environment (e.g., the impact of small infrastructure along the street) on the model.

\section*{Acknowledgment}
The work has been realized within the project no. 2018/29/B/ST7/01241 funded by the National Science Centre in Poland.

\ifCLASSOPTIONcaptionsoff
  \newpage
\fi



\bibliographystyle{IEEEtran}
\bibliography{bibliografia}
%


%

\begin{IEEEbiography}[{\includegraphics[width=1in,height=1.25in,clip,keepaspectratio]{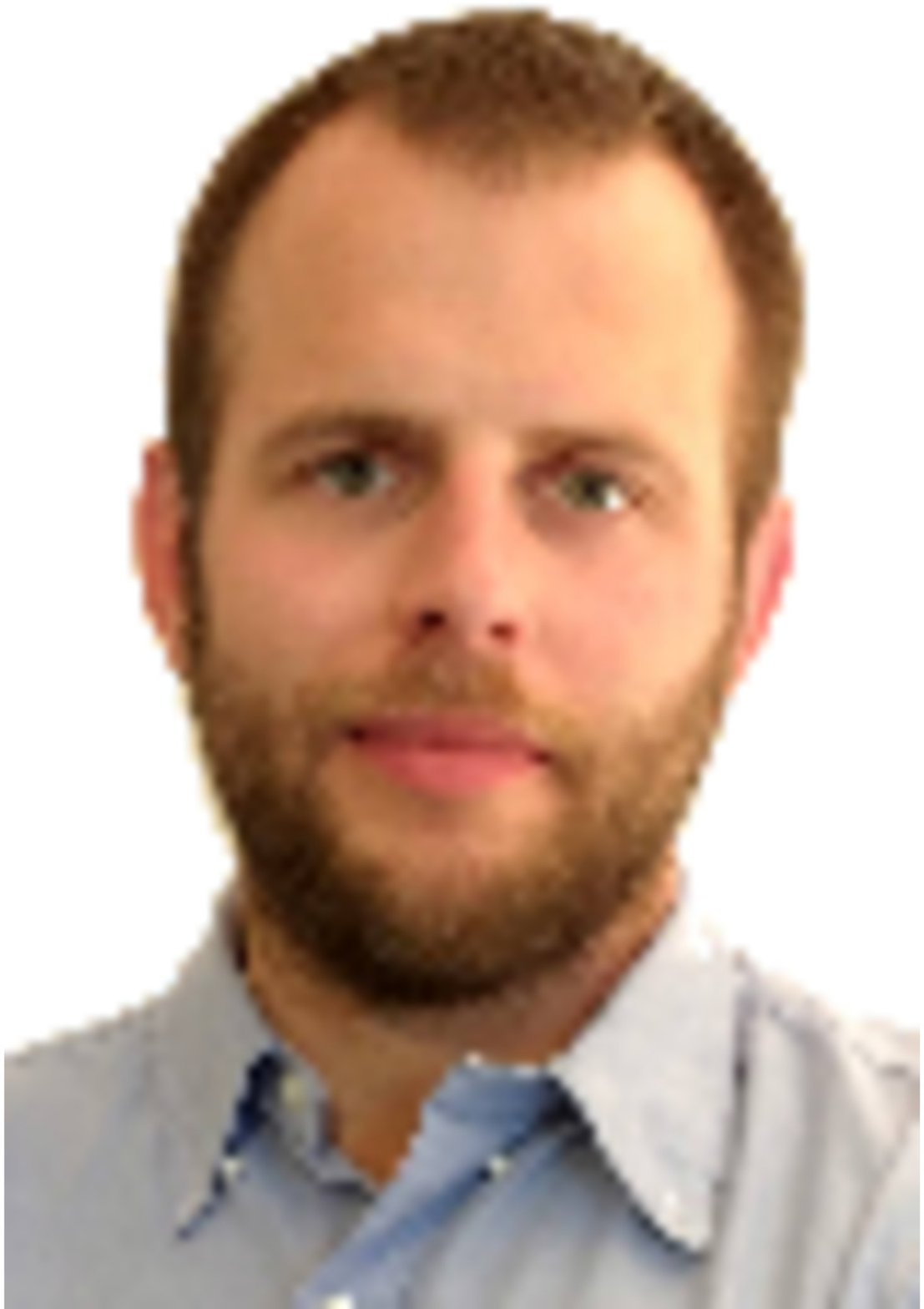}}]{Pawel Kryszkiewicz} received Ph.D. and postdoctoral degree in telecommunications from the Poznan University of Technology (PUT), Poland, in 2015 and 2022, respectively. He is currently an Assistant Professor with the Institute of Radiocommunciations, PUT. He has been involved in a number of national and international research projects. His main fields of interest are multicarrier system design, green communications, Dynamic Spectrum Access and Massive MIMO systems.
\end{IEEEbiography}

\begin{IEEEbiography}[{\includegraphics[width=1in,height=1.25in,clip,keepaspectratio]{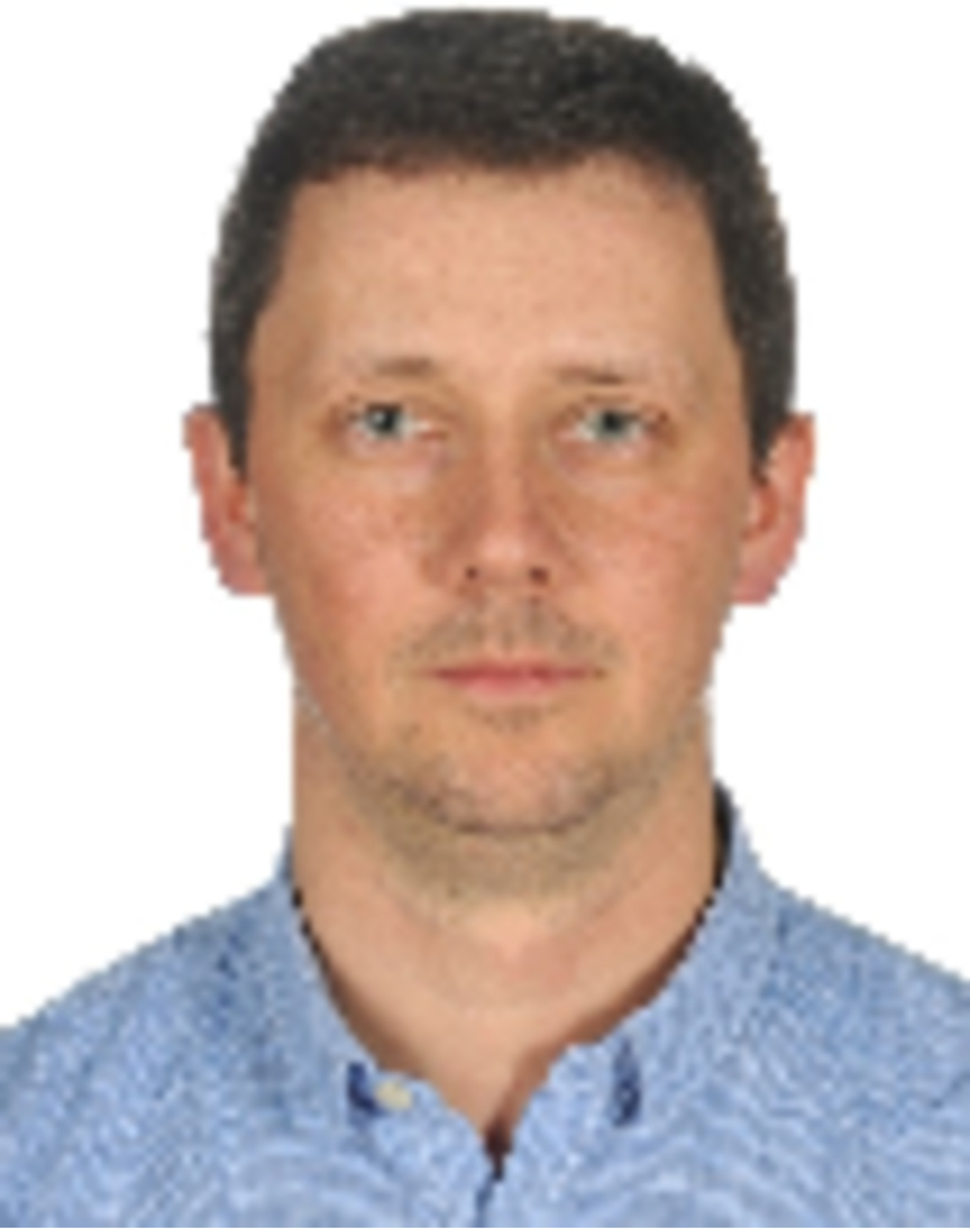}}]{Pawe\l{}~Sroka}
received the M.Sc. and Ph.D. (with honors) degrees in telecommunications from Poznan University of Technology (PUT), Poland, in 2004 and 2012, respectively. Currently he is employed as an Assistant Professor at the Institute of Radiocommunications, Faculty of Computing and Telecommunications, PUT. For several years he has participated in various international and national research projects.Paweł Sroka is an author of more than 40 scientific publications. His main research interests include 5G systems, radio resource management for wireless networks, vehicular communications (V2X), cross-layer optimization and MIMO systems.
\end{IEEEbiography}

\begin{IEEEbiography}[{\includegraphics[width=1in,height=1.25in,clip,keepaspectratio]{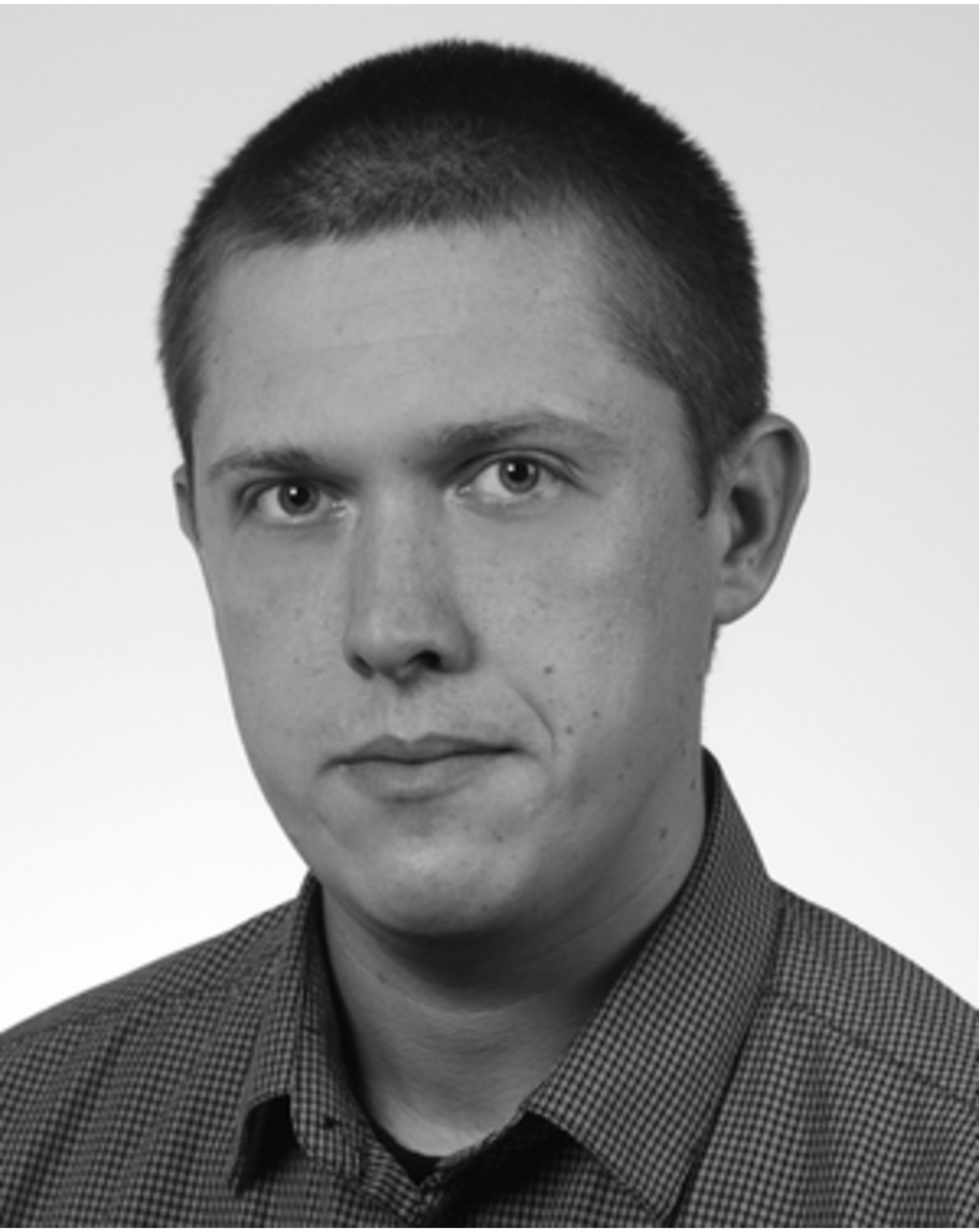}}]{Micha\l~Sybis} 
received the M.Sc. in electronics and telecommunications and Ph.D (with honors) degree in telecommunications from the Poznan University of Technology, Poznan, Poland in 2007 and 2012, respectively. He is currently an Assistant Professor with the Institute of Radiocommunications at the Faculty of Computing and Telecommunications at Poznan University of Technology. His research interests include coding techniques, iterative decoding, the fifth generation mobile wireless systems and vehicular communications.
\end{IEEEbiography}

\begin{IEEEbiography}[{\includegraphics[width=1in,height=1.25in,clip,keepaspectratio]{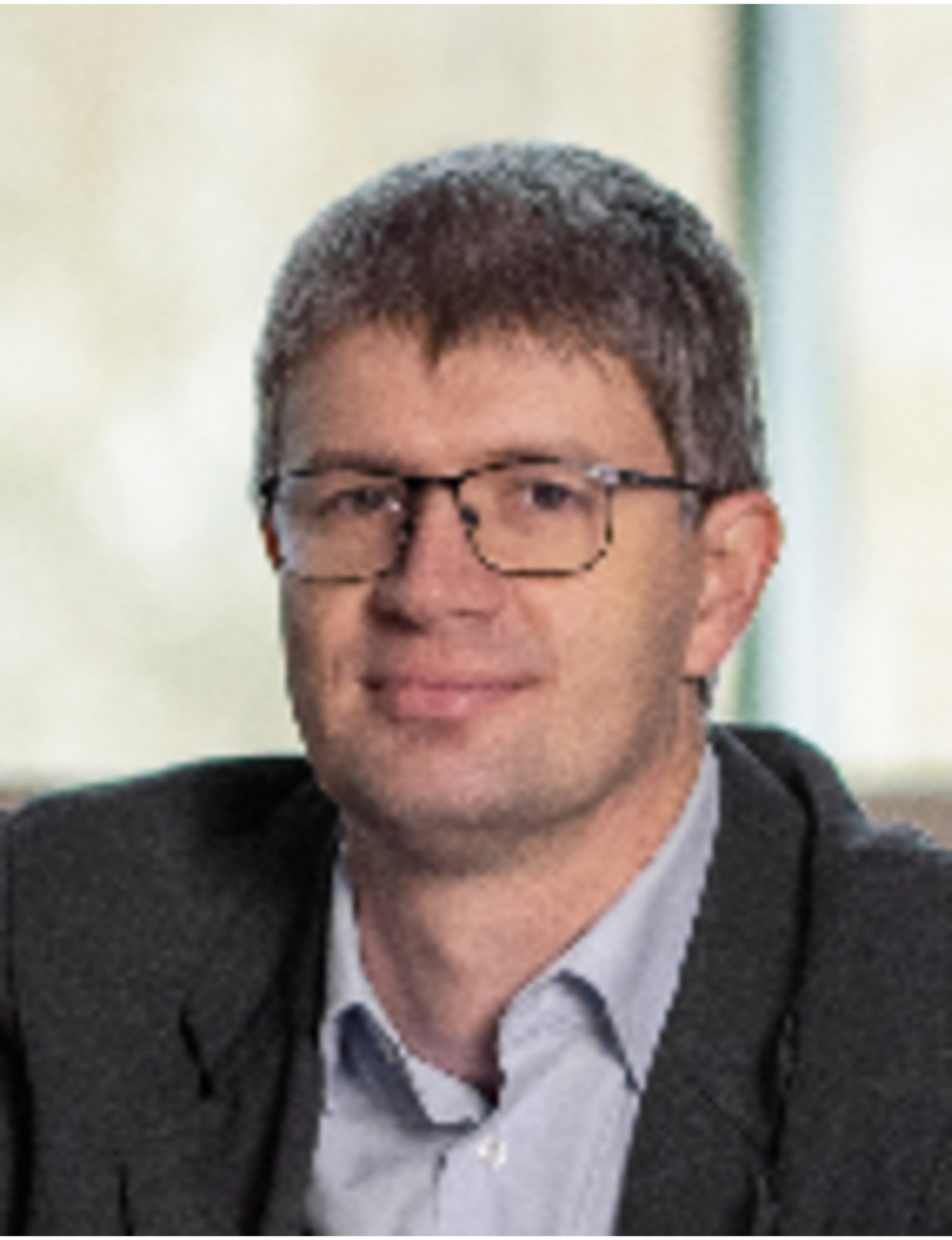}}]{Adrian Kliks} received his Postdoctoral degree in Technical Sciences, discipline: Technical Computer Science and Telecommunications in February 2019. He works as a University Professor at the Institute of Radiocommunications of the Poznan University of Technology. He took part in numerous international research projects: URANUS, NEWCOM++, ACROPOLIS, COGEU, NEWCOM\#, COHERENT,
in COST IC0902 and COST-Terra (IC 0905) and in national projects EcoNets, Bionets, OPUS project manager on V2X communication, and did he manage numerous industrial and commissioned projects. A member of the IEEE for many years, IEEE Senior Member since 2013, a member of the IEEE Broadcasting Society, IEEE Communication Society, IEEE Standard Association. Dr hab. Adrian Kliks, participated in the years 2012-2017 in the work of the IEEE P1900.x standardization group, as a member with voting rights, and also served as secretary. Member of the groups: Radio Communications Committee and Research Group on Software Defined and Virtualized Wireless Access. In the years 2014-2016 he was the Membership Development / Web Visibility Chair in the IEEE for the EMEA area. From 2019 – editor-in-chief of the Journal of Telecommunications and Information Technology of the Institute of Communications
\end{IEEEbiography}







\end{document}